\documentclass[12pt]{revtex4-1}
\usepackage{epsfig}
\usepackage{amssymb}
\usepackage{amsmath}
\usepackage{subcaption}
\usepackage{verbatim}
\usepackage{algorithm}
\usepackage{algorithmic}
\newcommand\opround{\operatorname{round}}
\newcommand\deltaf{\mathit{{\delta}{f}}}

\expandafter\let\csname equation*\endcsname\relax
\expandafter\let\csname endequation*\endcsname\relax

\newcommand{\romanNum}[1]{\uppercase\expandafter{\romannumeral#1}}

\usepackage{graphicx}
\usepackage{setspace}
\usepackage{mathrsfs}
\usepackage[table]{xcolor}
\usepackage{hyperref}
\hypersetup{colorlinks, citecolor=black, filecolor=black, linkcolor=black, urlcolor=black}

\begin{document}
\author{B.F. McMillan}
\affiliation{Centre for Fusion, Space and Astrophysics, Department of Physics, University of Warwick, CV4 7AL, Coventry UK}
\author{J. Ball}
\affiliation{Ecole Polytechnique F\'{e}d\'{e}rale de Lausanne (EPFL), Swiss Plasma Center (SPC), CH-1015 Lausanne, Switzerland}
\author{S. Brunner}
\affiliation{Ecole Polytechnique F\'{e}d\'{e}rale de Lausanne (EPFL), Swiss Plasma Center (SPC), CH-1015 Lausanne, Switzerland}
\date{\today}
\title{Simulating background shear flow in local gyrokinetic simulations}

\begin{abstract}
  Local gyrokinetic simulations solve the gyrokinetic equations with homogeneous background gradients, typically using a doubly periodic domain in the $(x,y)$ plane (i.e. perpendicular to the field line).  Spatial Fourier representations are almost universal in local gyrokinetic codes,
  and the {\it wavevector-remap} method was introduced in [Hammett et. al., Bull Am Phys Soc {\bf VP1} 136, (2006)] as a simple method for expressing the local
  gyrokinetic equations with a background shear flow in a Fourier representation. Although extensively applied, the wavevector-remap method has not been formally
  shown to converge, and suffers from known unphysicality when the solutions are plotted in real space [Fox et. al. PPCF {\bf 59}, 044008]. In this work, we use an analytic solution in slab geometry to demonstrate that wavevector-remap leads to incorrect {\it smeared} non-linear coupling between modes. We derive a correct, relatively simple method for solving local gyrokinetics in Fourier space with a background shear flow, and compare this to the
  wavevector-remap method. This allows us to show that the error in wavevector-remap can be seen as an incorrect rounding in wavenumber
  space in the nonlinear term. By making minor modifications to the nonlinear term, we implement the corrected wavevector-remap scheme in the GENE\cite{GENE} code and compare results of the original and corrected wavevector-remap for standard nonlinear benchmark cases. Certain physical phenomena are impacted by the errors in the original remap scheme, and these numerical artefacts do not reduce as system size increases: that is, original wavevector-remap scheme does not converge to the correct result.  
\end{abstract}

\maketitle

The numerical questions this paper deals with, although presented in the framework of gyrokinetic theory, are essentially about the generic question of how to model advection-type equations for periodic problems with a homogeneously sheared background flow. Such flows arise in various physical scenarios including astrophysical\cite{Umurhan2004} and fusion\cite{CooperPPCF30,RitzRotShearTurbSuppression1990, BurrellShearTurbStabilization1997} plasmas. When there is a background shear flow, the unit cell of this periodic turbulence must either shear with the background flow, or have shearing-box boundary conditions, to be consistent with the evolution of the initial periodic state. The underlying question is how to correctly represent shearing-box boundary conditions when using a spectral representation of the advecting fields. We will present this in the context of gyrokinetic theory\cite{Hahm_1988}, which solves for the evolution of the distribution of charged particles in a plasma as they interact with electromagnetic fields (although various extensions also exist), with typical variation timescales longer than the particle gyration time.
  
The wavevector-remap method, as introduced in ref. \cite{HammettRemap}, was intended to solve the local gyrokinetic equations in
a periodic flux tube with a homogeneously sheared background flow (we call this method {\it original wavevector-remap}). The original treatment interprets it
as a simplification of the moving-grid method\cite{HammettRemap}, where the wavenumber grid becomes increasingly sheared as time proceeds. This is analogous to treating the system of equations in the Lagrangian frame moving with the background fluid velocity.
Wavevector-remap is given an intuitive justification in terms of combining the standard evolution in the absence of a shear flow, and adding the evolution due to the background shear flow using a discrete wavevector shifting operation: instead of a smooth advection of complex wave amplitude through wavenumber space, the background shear leads to periodic discrete jumps. This is usually thought of as a kind of time-splitting operation, but the non-smoothness of the distribution functions in wavenumber space (the Fourier coefficients of smooth functions are generally not smooth) 
prevents the application of the usual proofs of convergence. Certain worrying unphysical features
of remap have been observed\cite{candy_remap}, especially in real space movies, where there are discontinuous jumps in the value of quantities\cite{Fox}.
We show that the wavevector-remap leads to an order $1$ error for a simple case where two plane waves couple
with each other nonlinearly. These difficulties with wavevector-remap have motivated other methods for simulating plasmas with sheared flows, such as the method presented in ref. \cite{candy_remap}, which uses radially periodic shear flows to avoid the need for shear-periodic boundary conditions.

To explain why the wavevector-remap method nonetheless gives correct results in certain cases, we consider a Fourier space method
derived using a moving-grid interpretation. This leads to a formally correct method for solving the local gyrokinetic equations.
By comparison with the correct method, we can show that the error in wavevector-remap results from an incorrect nonlinear coupling
term which is effectively a rounding error in wavenumber space.

We follow the procedure of deriving the gyrokinetic equations for an infinite set of modes on a grid moving smoothly through wavenumber space (the {\it moving grid method}), and then we choose a finite set of these modes to evolve numerically.
This discretisation involves a remap-like approach (which we call {\it corrected wavevector-remap}) to account for which modes are kept in a simulation, and as an indexing scheme for practical storage in computer memory. This separates
the question of which equations are being solved for each mode from the practical problem of how to implement and store these calculations in practise.

It is not the case that this moving-grid approach is highly expensive computationally: the linear evolution may be calculated
using the same methods of wavevector-remap. Previous claims that moving-grid method would be expensive\cite{CassonPHD,HammettRemap}
focused on the cost of accurately computing linear terms at each timestep to account for a continuously varying wavenumber. But
inaccuracy in the linear terms, which are smooth in wavenumber, has been shown to have little effect in certain limits\cite{CassonPHD}.
As we will discuss, the decision to effectively round the wavenumber to the nearest grid point, however, causes a lowest-order error in the nonlinear
computation so the wavevector-remap method is not a convergent numerical method in the formal sense (for particular problems it may however give reasonable results).

We show that the standard remap method leads to an unphysical smearing of nonlinear coupling with width of order the Fourier grid spacing.
It might be argued that the simulation will thus `converge' as Fourier space resolution is increased. This, however, is not in line with standard definitions
of the meaning of numerical convergence. Typically the pointwise error is used
to evaluate whether an approximate solution is close to the correct solution, and this does not converge to zero in the high-resolution limit. This is because in general the
distribution function $g$ is not smooth in wavenumber space, so arbitrarily displacing energy by one grid cell in Fourier space implies a very different form in real space.
Note that even though amplitudes of Fourier modes in a nonlinear simulation may look smooth, especially when averaged over time and/or plotted on a logarithmic scale, mode phases are essentially uncorrelated between neighbouring modes. 

With the standard remap method, the region of Fourier space where nonlinear coupling occurs is approximately correct, so some aspects of the nonlinear process are nonetheless preserved, and it may thus be that certain weak-turbulence results can still be reproduced. However, it seems simpler and more satisfying to use a provably correct method to solve for gyrokinetic turbulence in the presence of a background flow than attempt to estimate in much more detail how these errors might affect standard wavevector-remap simulations: direct comparison with corrected remap simulations will then allow us to determine whether previous results are in error. 

Especially to researchers who perform most of their analysis in spectral space, rather than examining real space quantities, it might not seem obvious that
small errors in wavenumber  associated with the original remap scheme (that we will later quantify) are problematic. For example, taking a 2D plot of mode amplitudes (conventionally a density plot using a logarithmic scale for mode amplitudes) in a well-resolved simulation and shifting quantities by one pixel would not make a dramatic difference to the figure. We highlight, however, using a visual example, that it makes a dramatic difference to structures in real space. A sharp spatial structure like a step function must be represented using a large number of plane wave components with specific phases chosen to allow constructive and destructive interference in the appropriate locations.

\begin{figure}
\begin{subfigure}{0.48\textwidth}
\includegraphics[width=8.5cm]{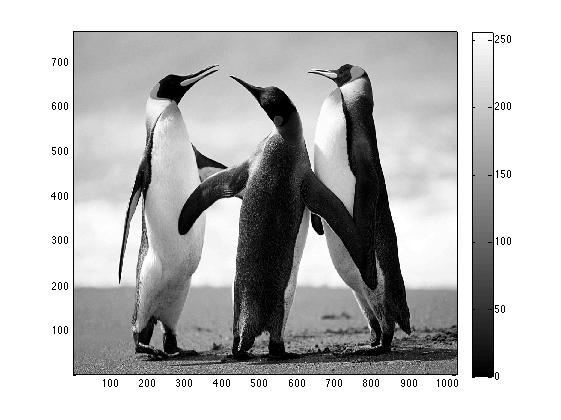} \caption{} \label{fig:penguins}
\end{subfigure}
\begin{subfigure}{0.48\textwidth}
\includegraphics[width=8.5cm]{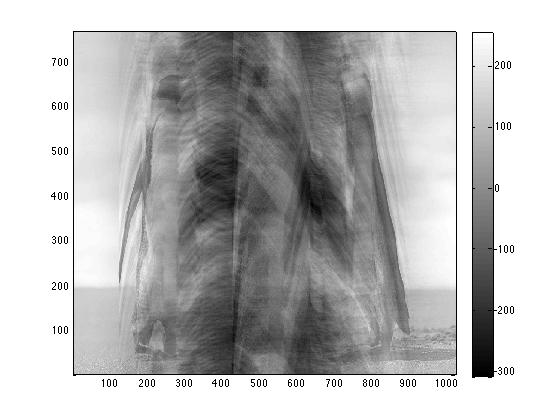} \caption{} \label{fig:penguins_remap}
\end{subfigure}

\caption{ (a) A grayscale digital photographic image\cite{penguin} (b) subjected to a one-unit shifts in horizontal wavenumber designed to simulate errors in the remap operation.} 
\label{fig:penguins_remapped}
\end{figure}

We use a greyscale digital photographic image (figure \ref{fig:penguins_remapped}) of a scene containing multiple penguins to illustrate how the errors in remap might have an impact on spatial structures. After taking the two-dimensional FFT to produce image mode amplitudes $\phi_{I,J}$ (with horizontal Fourier index $I$ and vertical index $J$), we perform the transform $\phi_{I+K(J),J} \rightarrow \phi_{I,J}$, with a randomly chosen $K(J) \in \{-1,0,1\}$. Note that the $I$ index is treated using modulo arithmetic, and the reality condition is applied by using $K(-J)=-K(J)$. This procedure simulates the off-by-one error that arises in the original remap procedure. After taking the inverse FFT of $\Phi$, we produce an image illustrating the effects of this single-index shifting. Near the left and right sides of the domain, there is little modification to the image, but in the centre of the domain, sharp features are broken up and the original structures are largely unrecognisable. 

\section{The moving-grid method}

To understand why the wavevector-remap method leads to correct linear simulations in the limit of large $k_x$ resolution (it would otherwise have been immediately dismissed), but
the details of nonlinear coupling are incorrect,
we follow a shear-periodic description of the problem, which is conceptually equivalent to the moving-grid description of \citet{HammettRemap}.
First, we will derive a convergent approximation for gyrokinetics with a background shear flow, then compare it to the remap method.

The gyrokinetic Vlasov equation for the perturbed distribution function $\delta f$ in the local limit\cite{GENE,BeerBallooningCoordinates1995} may be written symbolically as
\begin{equation}
  0 = \partial_t \deltaf + [ \langle \Phi \rangle, \deltaf] + L_1(\mu,v_{||},z) \deltaf + [\langle \Phi \rangle, f_0] ,
  \label{eq:gksymb}
\end{equation}
where $L_1$ is the linear gyrokinetic operator accounting for background drifts
and parallel motion and acceleration, $\langle\Phi\rangle$ is the gyroaveraged electrostatic field, $f_0$ is a background distribution function, and $\mu$ and $v_{||}$ 
are the magnetic moment and parallel velocity. The square brackets are Poisson brackets representing advection due to the $E \times B$ velocity implied by the electrostatic field. We have 
\begin{equation}
   [f,g] = \nabla f \times \nabla g . \frac{\mathbf{\hat{b}}}{B} = \frac{1}{C} \left[ \frac{\partial f}{\partial x} \frac{\partial g}{\partial y} + \frac{\partial f}{\partial y} \frac{\partial g}{\partial x} \right]
\end{equation}
where the magnetic field unit vector $\mathbf{b}$ and strength $B$ appear.
 The spatial coordinates parameterise the position $\mathbf{R}(x,y,z)$ with $x$ and $y$ labelling the two directions perpendicular to the background magnetic field and $z$ varying along the field line. 
$y$ is the binormal coordinate, with $d\mathbf{R}/dy$ in the toroidal direction. In this coordinate system, the background magnetic field may be written $\mathbf{B} = C \nabla x \times \nabla y$, where $C$ is a constant.
The final Poisson bracket in eq \ref{eq:gksymb} is linear in $\Phi$ and only a function of $z$ and $\mu$.

A symbolic, rather than explicit notation, is being used because we are more interested in the question of how these terms behave as operators (such as their smoothness) rather than their explicit form. The issues to be understood are really related to the general problem of convection in a shearing-box periodic domain rather than specific to gyrokinetics.

Given a radially varying background electric field of the form $\Phi_0 =  C S x^2 /2$, which leads to a uniformly shear background $E \times B$ flow  in the $y$ direction,  
we define a perturbed field $\phi$ using $\Phi = \phi + \Phi_0$ and find
\begin{equation}
  0 = \partial_t \deltaf + [ \langle \phi \rangle, \deltaf] + x S \partial_y \deltaf + L_1(\mu,v_{||},z) \deltaf
                   + [\langle \phi \rangle, f_0] . \label{eq:gksymPerturbed}
\end{equation}

To simplify the discussion in the rest of the paper, the parallel boundary conditions along the field line are not discussed in detail -- they are handled the
same way in the various numerical methods described and thus are not of immediate relevance\cite{BeerBallooningCoordinates1995}. Similarly, electromagnetic fluctuations are not discussed.

Since $\phi$ is a linear function of $\deltaf$ through the Poisson equation, we can simplify the notation further to
\begin{equation}
  0 = \partial_t \deltaf + [ \langle\phi\rangle, \deltaf] + x S \partial_y \deltaf + L_2(\mu,v_{||},z) \deltaf
\end{equation}
where $L_2 = L_1 + [\langle \Phi \rangle,f_0]$ is an integro-differential operator, homogeneous in $x$ and $y$.

Consider the Fourier representation
\begin{equation}
   \deltaf(x,y,t) = \sum_{I,J = -\infty}^{\infty} \overline{\deltaf}(I,J,t) \exp\left[i x ( I k_{x0} - S J k_{y0} t) + i y J k_{y0} \right]
   \label{eq:fourierrep}
\end{equation}
for integer $I$ and $J$, where $k_{x0} = 2 \pi / L_x$,  $k_{y0} = 2 \pi / L_y$, $L_{x}$ and $L_{y}$ are the lengths of the shear-periodic domain in the $x$ and $y$ direction respectively, and the overbar denotes the
Fourier transform of a given quantity (note the reality condition implies $\overline{\deltaf}(I,J,t)=\overline{\deltaf}^*(-I,-J,t)$). For a certain mode $(I,J)$, we have 
\begin{equation}
    (k_x,k_y) = ( I k_{x0} - S J k_{y0} t, J k_{y0}). \label{eq:kvals}
\end{equation}
This Fourier representation satisfies the periodicity constraint $\deltaf(x,y,t) = \deltaf(x + A L_x, y + B L_y + A t L_x S)$ for integers $A$ and $B$,
so at $t=0$ a simple double periodicity in $x$ and $y$ is implied. At later time, the representation is shear-periodic: this is
necessary to be consistent with the uniform background shear flow.

The real space $\deltaf$ arising from $\overline{\deltaf}(I,J,t)= \overline{\deltaf}(I,J,0)$ is trivially a solution to the advection equation
\begin{equation}
   \partial_t \deltaf(x,y,t) = -x S \partial_y \deltaf
\end{equation}
as the time-shift in the Fourier representation (eq. \ref{eq:fourierrep}) represents the spatial advection process. 

Inserting the Fourier series representation of equation (\ref{eq:fourierrep}) into the gyrokinetic equation and separating each Fourier component leads to
\begin{equation}
  0 = \partial_t \overline{\deltaf} + W(\overline{\deltaf},\phi) + \bar{L}_2(k_x , k_y, \mu, v_{||},z) \overline{\deltaf} , 
  \label{eq:movinggrid_gyrok}
\end{equation}
where $W$ represents the nonlinear Poisson bracket. Note that advection by the background shear flow does not appear explicitly in eq. \ref{eq:movinggrid_gyrok} because it has been
accounted for by the form of the Fourier transform. The Fourier space representation of the linear term $L_2$, $\bar{L}_2$, does not couple Fourier
modes with different $\bf{k}$, due to the homogeneity of the system in $x$ and $y$ (except for the parallel boundary condition, which accounts for the linearised safety factor profile). The operator $W$ is the transform of the Poisson bracket $[\langle \phi \rangle,\deltaf]$ and its Fourier components may be defined as
\begin{equation}
  W(I,J) = L_x^{-1} L_y^{-1} \int dx dy [\langle\phi\rangle,\deltaf] \exp \left[-i x ( I k_{x0} - S J k_{y0} t) - i y J k_{y0} \right].
  \label{eq:fourierW}
\end{equation}


In real space (i.e. not Fourier space), the nonlinear term is given by 
\begin{align}
[\langle \phi \rangle, \deltaf]  = & \nabla \langle \phi \rangle \times \nabla \deltaf . \frac{\mathbf{\hat{b}}}{B} \notag \\
 = &-\left[ \sum_{I',J'} \langle \bar{\phi} \rangle (I',J',t) \mathbf{k}(I',J',t) \right] \times
    \left[ \sum_{I'',J''} \overline{\deltaf}(I'',J'',t)\mathbf{k}(I'',J'',t) \right] .\frac{\mathbf{\hat{b}}}{B} \nonumber \\
&\exp\left\{ i x (k'_x + k''_x - S [k'_y+k''_y] t) + i y (k'_y + k''_y) \right\} \label{eq:movinggrid_nonlin_unsimplified} \\
 = &-\sum_{I',J',I'',J''} \langle \bar{\phi} \rangle(I',J',t) \overline{\deltaf}(I'',J'',t) \mathbf{k(I',J',t)}\times\mathbf{k(I'',J'',t)}.\frac{\mathbf{\hat{b}}}{B} \nonumber \\
&\exp\left\{ i x (k'_x + k''_x - S [k'_y+k''_y] t) + i y (k'_y + k''_y) \right\}.
  \label{eq:movinggrid_nonlin}
\end{align}
where the wavevector $\mathbf{k}(I,J,t) = k_x \nabla x + k_y \nabla y$.

For practical numerical purposes, one can directly evaluate equation (\ref{eq:movinggrid_nonlin_unsimplified}) on a grid in real space (using the modified Fourier transform)
and transform back numerically into the (modified) Fourier space (as we will show later). However, for the moment we analytically
project equation (\ref{eq:movinggrid_nonlin}) into Fourier components (using equation (\ref{eq:fourierW})) and find
\begin{equation}
  W(I,J) = -L_x^{-1} L_y^{-1} B^{-1} \sum_{I',J',I'',J''} \delta_{I' + I'',I} \delta_{J' + J'',J}  \langle\bar{\phi}\rangle(I',J') \overline{\deltaf}(I'',J'') \mathbf{k}(I',J',t) \times\mathbf{k}(I'',J'',t).\mathbf{\hat{b}} , \label{eq:nonlinearTerm}
\end{equation}
which has the same form as the standard nonlinear term used in spectral local gyrokinetic computation. The only difference is that the wavenumbers $\mathbf{k}$
depend on time (the time-dependence does not enter the coupling condition). When explicitly evaluated, the time dependence also drops out of the wavenumber cross-product [ie $\mathbf{k}(I,J,t) \times \mathbf{k}(I,J,t) = \mathbf{k}(I,J,0) \times \mathbf{k}(I,J,0)$].
An equivalent way to calculate this more efficiently is via a Fourier convolution approach, i.e
\begin{equation}
   W_{I,J} = F \left( F^{-1}(\mathbf{A}) . F^{-1}(\mathbf{B}) \right) ,
\end{equation}
where $F$ is the standard discrete Fourier transform, $\mathbf{A}(I,J) = (\mathbf{\hat{b}}/B) \times \overline{\deltaf}(I,J) \mathbf{k}(I,J,t) $ and $\mathbf{B}(I,J) = \langle \bar{\phi}(I,J) \rangle \mathbf{k}(I,J,t)$. 

\begin{figure}
  \centering
  \includegraphics[width=\textwidth]{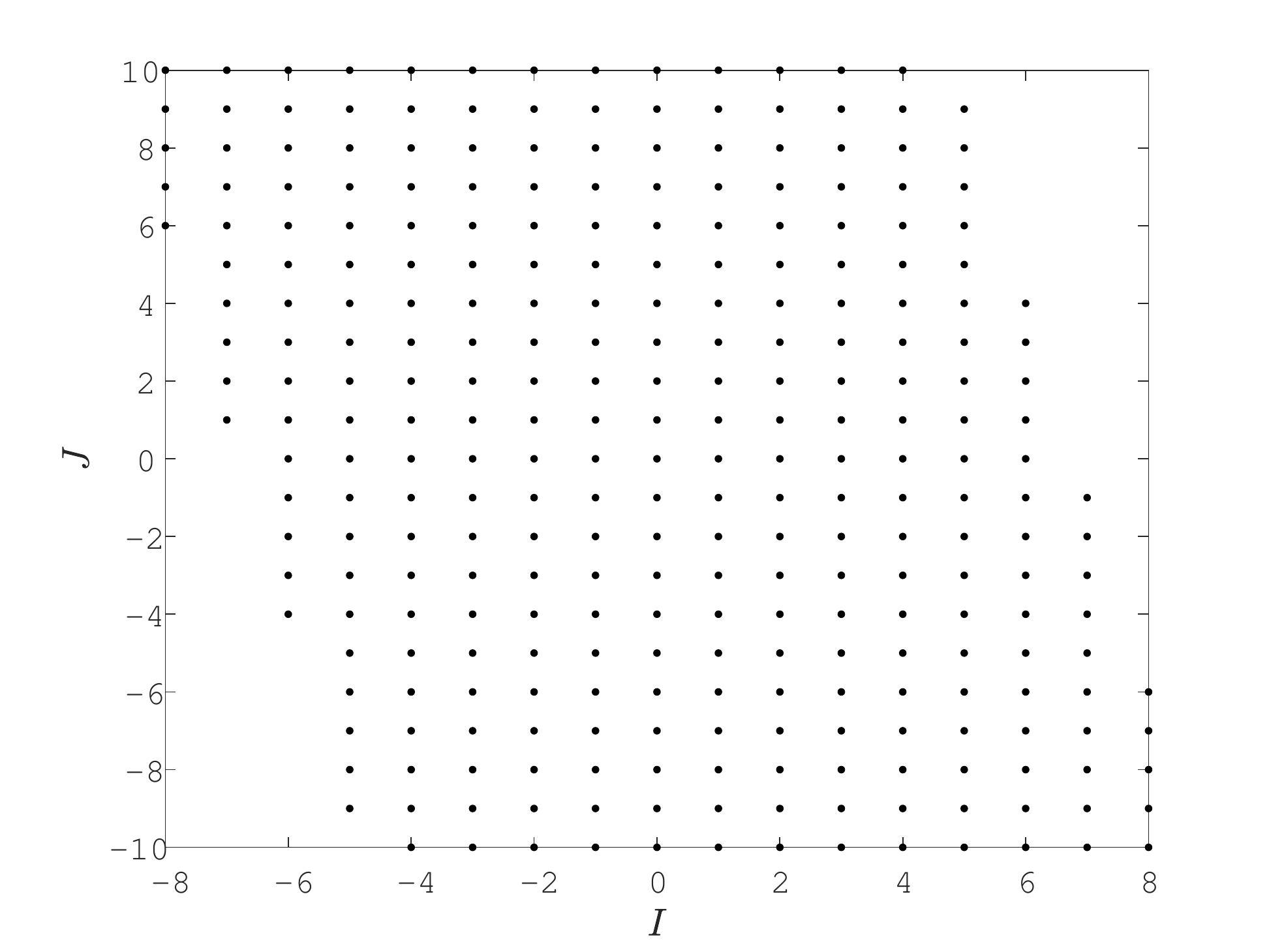}
  \caption{The set of modes kept in a moving grid description with $k_{x0}=k_{y0}=1$, and $S t = -0.2$ if all modes with $|k_{x}|<7$ and $|k_y|<11$ are retained.}
  \label{fig:shearedmodepic}
\end{figure}

The time evolution of $\overline{\deltaf}(I,J)$ can be written using equations (\ref{eq:movinggrid_gyrok}) and
(\ref{eq:movinggrid_nonlin}): this is a spectral representation of local gyrokinetics with a background equilibrium
shear flow and the appropriate shear-periodic boundary conditions.
The difference with the standard local gyrokinetic scheme (without background shear flow) is not very large, but there are two main
points of difference. First, the $x$ wavenumber on the moving-grid increases with time, but we can practically only solve
for a finite number of wavenumbers. In order to keep all the modes up to a certain wavenumber $K_x$,
we would evolve each of these modes on the moving-grid only during the time where $ |I k_{x0} - J S k_{y0} t| < K_x $.
Usually the
actual grid of modes solved will therefore not be a rectangular array in $(I,J)$ space, but rather the sheared domain shown in figure \ref{fig:shearedmodepic}, that secularly becomes more sheared at late time.

The second practical difference to standard gyrokinetic computation is that $\bar{L}_2$ is now evaluated at arbitrary $k_x$,
rather than on a grid. Let us now use the approximation
$k_x  \approx  \opround(k_x / k_{x0}) k_{x0}$ (where $\opround(s)$ is the integer closest to $s$) when evaluating $\bar{L}_2$,
so the evolution equation becomes
\begin{equation}
  0 = \partial_t \overline{\deltaf} + W(\overline{\deltaf},\phi) +  \bar{L}_2(k_{x0} \opround( k_x/ k_{x0}), k_y, \mu, v_{||},z) \overline{\deltaf}
  \label{eq:movinggridapprox}
\end{equation}
and $\bar{L}_2$ is now evaluated exclusively at points on the usual discrete Fourier grid $(I k_{x0} , J k_{y0})$. Note that $\bar{L}_2$ needs
to be relatively smooth for this approximation to be appropriate and in general small $k_{x0}$ is required for
good approximation (e.g. for capturing the variation of Bessel functions we require $k_{x0} \rho \ll 1$). In general, $\bar{L}_2$ is a smooth analytic function of $\mathbf{k}$, except possibly near $k=0$. It potentially has singular behaviour near $k=0$ because terms that involve $\phi$ (such as the Poisson bracket $[\langle\phi\rangle,f_0]$) may have, at long wavelength, a $k^{-2}$ dependence on $\deltaf$ through the long wavelength Poisson equation $(A - k^2) \phi \propto \int dv^3 \deltaf$. The coefficient $A$ is non-zero if there is an adiabatic response term (representing, for example an electron background). In cases where $A=0$, such as fully kinetic runs, the operator $L_2$ may be quite different for modes with wavenumber $(0,k_{y0})$ and $(k_{x0},k_{y0})$, in the typical case where $k_{x0} \sim k_{y0}$. Therefore, in general, the treatment of these box-scale modes may be somewhat inaccurate: however, even in the absence of background flow, simulations with substantial amplitudes in these box-scale modes are generally considered not to be converged. 


Let us
compare this approach to the original remap method. In the original remap method each of the
wavenumbers are indexed using the `nearest point' in the initial $t=0$ grid of wavenumbers.
That is, the mapping between the Fourier coefficients of the distribution $\bar{g}$ in the moving-grid method and $\widetilde{\deltaf}$,
the Fourier coefficients in the remap description, is
\begin{equation}
\overline{\deltaf}(I,J,t) = \widetilde{\deltaf}(I + \opround[J t/t_0],J,t)
\end{equation}
with $ t_0 =k_{x0}/ (S k_{y0})$.
In the linear regime where $W \rightarrow 0$ the approximate time evolution of equation (\ref{eq:movinggridapprox})
is now equivalent to the remap method once this mapping is performed.
In this spectral method, we expect the original wavevector-remap, in the linear regime, to be a correct approximation given a sufficiently small timestep and grid size in the $\mu$, $v_{||}$ and $z$ directions, presuming
the approximation to $\bar{L}_2$ is sufficiently good: good approximation for $\bar{L}_2$ is assured by choosing a sufficiently small $k_{x0}$. Note that we have not needed to make arguments here about `time-splitting'
schemes or concern ourselves about the smoothness of the distribution function in wavenumber space as in ref. \cite{HammettRemap}.
The reason statements about convergence can be made here in a relatively straightforward fashion is that the sudden jump in values of Fourier coefficients in the remap method
can equivalently be seen as just a relabelling of the grid index (to be dealt with later as an implementation detail)
and the smoothness of the overall evolution is much clearer in this point of view. The distribution function at least evolves continuously even though there are discontinuities in its rate of change (which are small for small $k_{x0}$).


The difference between the correct nonlinear coupling and that used in the original remap is now somewhat more clear. In the moving-grid
method the coupling is between modes with $ -I + I' + I'' = 0 $. On the other hand, translating indices $(I,J)$ to the corresponding
grid point in the remap method $(\hat{I},\hat{J})$ we have $\hat{I} = I + \opround(J t/t_0)$, and likewise for $I'$ and $I''$.
The nonlinear coupling in the remap
method is between $ - \hat{I} + \hat{I}' + \hat{I}'' = 0$,
and thus $ - I + I' +I'' - \opround(J t/t_0) + \opround(J' t/t_0) + \opround(J'' t/t_0) = 0$. Since we also have
$ - J + J' + J'' = 0$, we can rewrite this as
\begin{align}
  -I &+ I' +I'' -  [\opround(J t/t_0) - (J t/t_0)] \\
    &+ [\opround(J' t/t_0) - (J' t/t_0)] + [\opround(J'' t/t_0) - (J'' t/t_0)] = 0 ,  \nonumber
\end{align}
which is in general {\it not} identical to the moving-grid
method coupling condition of $ -I + I' + I'' = 0 $. Now each of the terms in square brackets has a magnitude of less than $1/2$ so the maximum error is $3/2$. On the other hand, we know that
$I+I'+I''$ must be an integer so the possible error in remap mode coupling is
\begin{equation}
   | I - I' - I'' | \le 1.
\end{equation}
In practise the absolute error is one grid point the majority of the time. 
This is basically the {\it rounding error} due to the arbitrary times at which modes are shifted in the remap method.
Note that there is an important special case when
this error is zero: if one of the modes interacting has $J=0$ (e.g. zonal modes), the nonlinear coupling is correct. Since
interactions with zonal  modes are very important to gyrokinetic simulation dynamics, this helps to explain why simulations using the original wavevector-remap method appear to give reasonable results in many situations.

As an example of where errors in the nonlinear term lead to qualitatively different behaviour, consider the cross product in the nonlinear coupling term. In the moving-grid method this cross-product is independent of time, but the original wavevector-remap a time dependence fictitiously arises as mode energy discretely shifts in radial wavenumber. Two modes with initially parallel wavenumbers, with zero cross product, can give rise to nonlinear coupling in original-remap, but this is exactly zero in the moving grid method.

The reason that off-by-one errors in the index leads to a large error here, but not for $L_2$
is that $g$ and analogously $\bar{g}$ (and therefore the nonlinear term as a whole) are not smooth functions in wavenumber space.
To put it another way, throwing away the fractional part of the normalised wavenumber $k_x/k_{x0}$ makes little difference
to the linear dynamics, but means that the wrong modes often couple nonlinearly, because rounding the sum of two numbers does
not necessarily give the same result as summing two rounded numbers. 
 
Demonstrating the error in the original wavevector-remap also shows what is required to correct a gyrokinetic code that uses it. The linear dynamics are now conceptually considered to be in a moving-grid in wavenumber
space, but the actual linear computation does not need to be modified. It is only the nonlinear calculation that requires modification.
Specifically, all that is required is to translate the indexes $(I,J) = (\hat{I} - \opround(\hat{J} t/t_0),\hat{J})$ and calculate the time-dependent $\mathbf{k}(I,J,t)$
values (using eq. \ref{eq:kvals}) that appear in the nonlinear terms. Using the exact values of $\mathbf{k}$ means that modes that are aligned initially (i.e. the cross product is initially zero) will remain so for all time. The re-indexing in the implementation of the original remap scheme already takes care of ensuring the
physically relevant range of $k_x$ are considered even in a moving-grid interpretation. 
Note that the moving-grid description has a time periodicity over $t_0$, so it periodically becomes aligned with the fixed rectangular grid.

\section{The modified Fourier transform and an alternative nonlinear evaluation}

Correctly evaluating real space quantities in a remap simulation
requires taking into account the time-evolving wavevectors, both for simulation diagnostics and for internal calculations. Since the nonlinear term is
evaluated in an original wavevector-remap simulation using a standard Fourier transform, we can see the difference between the original remap and the
moving-grid approach in terms of an incorrect evaluation of the nonlinear term in real space, followed by an incorrect transform of the
nonlinear term back into Fourier space.

The transform in equation (\ref{eq:fourierrep}) may be rewritten to enable use of standard Fourier transforms (and typical FFT routines) as
\begin{align}
  h(x,y,t) =&  \sum_{I,J} \bar{h}(I,J,t) \exp\left\{i x ( I k_{x0} - S J k_{y0} t) + i y J k_{y0} \right\} \\
  =& \sum_J \exp\left\{ i y J k_{y0} - i x S J k_{y0} t) \right\} \sum_I \bar{h}(I,J,t) \exp\left\{i x I k_{x0} \right\} \\
  =& \sum_J \exp\left\{ i y J k_{y0} \right\} \exp\left\{ - i x S J k_{y0} t \right\} \bar{h}_J(x,t) ,
   \label{eq:fourierep2}
\end{align}
where $\bar{h}_J \equiv \sum_I \bar{h}(I,J,t) \exp\left\{i x I k_{x0} \right\}$ is the result of a standard 1-D (inverse) Fourier transform over the $x$ direction. The last expression is now a standard 1-D Fourier transform
over the $y$ direction, but an important phase factor appears and must be included before the transform
is taken in the $y$ direction. The analogous back transform is simply the inverse of these steps, in the inverse order. Note that in general a non-rectangular grid of modes in $(I,J)$ space is stored so practical implementations of this would need some care.

The expression in eq \ref{eq:fourierep2} also leads to a definition of the real space field associated with a quantity $h$ stored as Fourier coefficients in the remap convention, as
\begin{align}
  h(x,y,t) =& \sum_{I,J} \tilde{h}(I,J,t) \exp\left\{i x (I k_{x0} - S J k_{y0} \delta t) + i y J k_{y0} \right\} \\
  =& \sum_J \exp\left\{ i y J k_{y0} \right \} \exp\left\{ - i x S J k_{y0} \delta t) \right\} \tilde{h}_J(x,t) \label{eq:correctedFourierTrans}
\end{align}
with, analogously, $\tilde{h}_J(x,t) \equiv \sum_I \tilde{h}(I,J,t) \exp\left\{i x I k_{x0} \right\}$
and where $\delta t = t - (t_0/J) \opround(J t/t_0)$ is related to the time since the last remap step for mode $J$. In these expressions, because a rectangular grid of modes is stored, we have $I \in [0,N_I-1]$ and $J \in [0,N_J-1]$ so the one-dimensional FFTs are entirely conventional.

This modified 2D Fourier transform leads to spatial quantities that are shear-periodic in the
$x$ direction (an alternative is output onto a time-varying grid that becomes increasingly sheared).
We have chosen $x=0$ to be the point where the background flow is stationary although this is of course arbitrary (some codes, e.g. GS2\cite{GS2_code}
take the centre of the domain to have zero flow). The additional phase factors are zero at $x=0$ so the real space quantities associated with the corrected- and original-remap method agree here. At $x=L_x$, the integer parts of the phase factor lead to a phase shift which is an integer multiple of $2 \pi$, so the factors can be written $\exp(-i L_x S J k_{y0} t)$, and the real-space quantities in the corrected remap are just translated in the $y$ direction by $S L_x t$ compared to the original remap. However, elsewhere in the $x$ domain the difference between these descriptions cannot be seen as a spatial translation, as the relative phases of modes do not rotate proportional to the $y$ wavenumber. The absence of these phase factors leads to `glitches' in the centre of the $x$ domain in movies made of real-space quantities in conventional remap method simulations; these glitches are due to the time-discontinuities in Fourier coefficients as they are `remapped', but in corrected-remap simulations jumps in the phase factors compensates for jumps in the coefficient values.

Since we have a straightforward method to calculate the real space quantities correctly from remap indexes, this provides a simple method to
modify a remap code to correctly evaluate the nonlinear term in gyrokinetic simulations. The correct nonlinear term is found from transforming
the continuous nonlinear terms $[\langle \phi \rangle,g]$, but an exact evaluation can also be made by evaluating the nonlinearity
on a fine grid (to avoid mode aliasing) and using a discrete transform. Let us evaluate the quantities entering the nonlinear term
on a fine grid in real space $X_m,Y_n$ with $ m \in [0 ,N_I^\dagger-1]$ and $n \in [0 , N_J^\dagger-1]$, and $(X_m,Y_n) = (m L_x/ N_I^\dagger, n L_y/ N_J^\dagger)$ (we will soon discuss how big $N_I^\dagger$ and $N_J^\dagger$ need to be). This is done, in practise, by first loading the relevant coefficients into the extended Fourier space of size $(N_I^\dagger, N_J^\dagger)$, which is larger than the Fourier grid used elsewhere in the code, and then applying the modified discrete Fourier transform (i.e. equation (\ref{eq:fourierep2}) evaluated on a grid).
One then multiplies the relevant terms together to evaluate the real space nonlinearity. We then transform back to Fourier space. The nonlinearity will
produce new terms outside the original Fourier grid: if we had not used an extended grid these short-wavelength terms would be {\it aliased}
to unphysically perturb longer wavelength modes within the original grid. Generally these short wavelength modes arising from nonlinear interaction are simply ignored.

Using the same logic as for calculating the rounding error, the possible interactions are for modes with $|I - I' - I''| \le 1 $ so the largest wavelength mode generated nonlinearly given the modes that lie within the remap grid of wavenumbers is $I = 2 N_I - 1$ instead of $2 N_I-2$ for standard spectral methods. The means the aliasing problem is slightly more severe. For standard remap methods, it is sufficient to choose a fine grid of size $(N_I^\dagger,N_J^\dagger)=(3/2) (N_I,N_J)$ (the {\it 3/2 rule}) but for the corrected-remap method we need one more grid cell in the $I$ direction, ie $N_I^\dagger= 3 N_I /2  + 1, N_J^\dagger = 3 N_J /2$.

In summary, the only modifications required to a standard remap code are to calculate the phase factor in the 1D FFTs
appearing in the nonlinear term (i.e. equation (\ref{eq:correctedFourierTrans})) for both forward and backward transforms, and to calculate the time-dependent wavenumbers appearing in the nonlinear term (eq. \ref{eq:nonlinearTerm}). It may also be necessary to slightly increase the number of real space $x$ grid points used to calculate nonlinear terms (to prevent aliased modes).

\section{Numerical implementation of the corrected wavevector-remap scheme}

The corrected wavevector-remap flow shear model (outlined in section II) has been implemented in the local version of the gyrokinetic turbulence code GENE \cite{JenkoGENE2000, GoerlerGENE2011}. GENE makes for a good example because, like most other local gyrokinetic codes, it uses a Fourier representation in the directions perpendicular to the magnetic field and previously used the original wavevector-remap scheme\cite{HammettRemap} described in section I. Implementing the corrected wavevector-remap method in such a code requires relatively little modification, all of which occurs in the calculation of the nonlinear term. As discussed in section II, the linear terms should {\it not} need to be modified as any error will converge to zero as the radial box size is increased\cite{ChristenFlowShear2018}. The $k_{x}$ grid spacing is inversely proportional to the radial box size, so a convergence study taking $k_{x0} \rightarrow 0$ will reveal if the error in the linear terms is significant (as is done in figure \ref{fig:amplitude}).

While nearly all local gyrokinetic codes are spectral, they typically calculate the nonlinear term in real space. This is done because it produces the same answer as a Fourier space calculation and is more computationally efficient. GENE treats the nonlinear term by first calculating the gyroaveraged turbulent $\vec{E} \times \vec{B}$ velocity in Fourier space according to
\begin{align}
  \vec{v}_{E} = - i \vec{k}_{\perp} J_{0} \left( k_{\perp} \rho_{s} \right) \overline{\phi} \times \frac{\hat{b}}{B} , \label{eq:FourierExBvelocity}
\end{align}
where we have assumed the turbulence is electrostatic for notational simplicity. Here $i$ is the imaginary unit, $\vec{k}_{\perp}$ is the perpendicular wavevector component, $J_{0}$ is the Bessel function of the first kind, $\rho_{s}$ is the gyroradius of species $s$, $\phi$ is the turbulent electrostatic potential, $B$ is the magnitude of the magnetic field, and $\hat{b}$ is its unit direction. GENE then calculates the gradient of the perturbed distribution function in Fourier space according to
\begin{align}
  \vec{\nabla} g_{s} = - i \vec{k}_{\perp} \overline{g}_{s} . \label{eq:FourierDistFnGradient}
\end{align}
At this point, instead of performing a convolution of the two quantities in Fourier space, they are transformed into real space, multiplied together, and the result is transformed back into Fourier space. 

As discussed previously, the key issue in the original wavevector-remap scheme is that $\overline{\phi}$ and $\overline{g}_{s}$ undergo discrete remaps in $k_{x}$ that happen at different times depending on the value of $k_{y}$. This means that modes do not stay in proper alignment and experience off-by-one errors. Since $\overline{\phi}$ and $\overline{g}_{s}$ are turbulent and not continuous in Fourier space, such off-by-one errors can be significant.

In order to eliminate off-by-one errors in the turbulent quantities, we must ensure that the modes stay exactly aligned. A simple and practical way to accomplish this is to keep the wavevector remap, but make the substitution $\vec{k}_{\perp} \rightarrow \vec{k}_{\perp} - S J k_{y0} \delta t \hat{e}_{x}$. Here $\delta t \equiv t - (t_{0} / J) \text{round} \left( J t / t_{0} \right)$ is related to the time since the mode was last remapped and $t_{0} \equiv k_{x0} / \left( S k_{y0} \right)$. This definition implies that $\delta t \in \left[ - t_{0}/(2J), t_{0}/(2J) \right)$ and $\delta t = 0$ when the mode is at a grid point. This modification is identical to a linear interpolation of the wavenumber between the two remaps and makes the wavenumber shift smooth and continuous. Note that such a linear interpolation is {\it not} an approximation; it is exact because the translation of the $k_{x}$ is indeed linear in time. This substitution must be made for the factor of $\vec{k}_{\perp}$ that appears in both equation (\ref{eq:FourierExBvelocity}) and equation (\ref{eq:FourierDistFnGradient}). 

Additionally, this substitution must be made in the Fourier transforms to and from real space. This process also has a dependence on the wavenumber vector, though it is hidden away in the discrete Fourier transform libraries called by the code. As discussed in deriving equation (20), when $\vec{k}_{\perp}$ is replaced with $\vec{k}_{\perp} - S J k_{y0} \delta t \hat{e}_{x}$ it leads a modified form of the Fourier transform:
\begin{align}
  h \left( x,y,t \right) = \sum_{J} \exp \left\{ i y J k_{y0} \right\} \exp \left\{ -i x S J k_{y0} \delta t \right\} H_{J} \left( x,t \right) .
\end{align}
Specifically, it introduces a phase factor of $\text{exp} \left( -i x S J k_{y0} \delta t \right)$. This factor adjusts for the fact that the actual $k_{x}$ value is often between the grid points that are being used in the discrete Fourier transform.

To accomplish this all in practice, we take
\begin{align}
  \vec{v}_{E} = - i \left( \vec{k}_{\perp} - S J k_{y0} \delta t \hat{e}_{x} \right) J_{0} \left( k_{\perp} \rho_{s} \right) \overline{\phi} \times \frac{\hat{b}}{B}
\end{align}
and
\begin{align}
  \vec{\nabla} \overline{g}_{s} = - i \left( \vec{k}_{\perp} - S J k_{y0} \delta t \hat{e}_{x} \right) \overline{g}_{s} ,
\end{align}
which are Fourier space quantities, and perform a standard discrete $k_{x} \rightarrow x$ Fourier transform on each. Then, we multiply each of the results by $\text{exp} \left( - i x S J k_{y0} \delta t \right)$, taking care to construct the $x$ grid to be consistent with the conventions of the discrete Fourier transform. In GENE, the following $x$ grid was used
\begin{align}
  X_{m} = m \frac{L_{x}}{N_{I}^\dagger} , \label{eq:xGrid}
\end{align}
where $m \in [ 0, N_{I}^\dagger-1 ]$ is the integer that serves as the grid index, $L_{x}$ is the width of the box in the $x$ direction, and $N_{I}^\dagger$ is the number of $x$ grid points. Note that, to prevent aliasing, GENE uses a value of $N_{I}^\dagger = (3/2) (N_{I} + \text{mod}(N_{I}, 2))$, where $N_{I}$ is the number of $k_{x}$ grid points and the $\text{mod}$ function gives the remainder of the first argument divided by the second. Additionally, for an even number of $k_{x}$ grid points, GENE permanently sets the highest grid point to 0. Thus, the preexisting value of $N_{I}^\dagger$ is sufficient to prevent aliasing and there is no need to add an extra grid point. After multiplying by the phase factor, a standard $k_{y} \rightarrow y$ Fourier transform is applied. Next, the $\vec{E} \times \vec{B}$ velocity and gradient of the distribution function, now real space quantities, are multiplied to produce the complete nonlinear term in real space. At this point, the process is reversed. A $y \rightarrow k_{y}$ Fourier transform is applied to the real space nonlinear term. Then, it is multiplied by $\text{exp} \left( + i x S J k_{y0} \delta t \right)$ and, finally, Fourier transformed in the $x \rightarrow k_{x}$ direction. Thus, the overall algorithm can be summarized in pseudo-code as written in algorithm \ref{algo:fft}, where the red text highlights the differences with the original wavevector-remap scheme. Note that the original scheme can be recovered by setting $\delta t = 0$.

 \begin{figure}
\begin{algorithm}[H]
\caption{Calculate $\overline{\left[ \left\langle \phi \right\rangle, g_{s} \right]}$, the nonlinear term in Fourier space}
\label{algo:fft}
\begin{algorithmic}
\STATE $\overline{v}_{E x} = i J k_{y0} J_{0} \left( k_{\perp} \rho_{s} \right) \overline{\phi}$
\STATE $\overline{v}_{E y} = i \left( I k_{x0} \textcolor{red}{- S J k_{y0} \delta t} \right) J_{0} \left( k_{\perp} \rho_{s} \right) \overline{\phi}$
\STATE $\overline{\partial g / \partial x} = i \left( I k_{x0} \textcolor{red}{- S J k_{y0} \delta t} \right) \overline{g}_{s}$
\STATE $\overline{\partial g / \partial y} = i J k_{y0} \overline{g}_{s}$
\STATE
\STATE $v_{E x} = F^{-1}_{k_{y} \rightarrow y} \left( F^{-1}_{k_{x} \rightarrow x} \left( \overline{v}_{E x} \right) ~ \textcolor{red}{\text{exp} \left( - i x S J k_{y0} \delta t \right)} \right)$
\STATE $v_{E y} = F^{-1}_{k_{y} \rightarrow y} \left( F^{-1}_{k_{x} \rightarrow x} \left( \overline{v}_{E y} \right) ~ \textcolor{red}{\text{exp} \left( - i x S J k_{y0} \delta t \right)} \right)$
\STATE $\partial g / \partial x = F^{-1}_{k_{y} \rightarrow y} \left( F^{-1}_{k_{x} \rightarrow x} \left( \overline{\partial g / \partial x} \right) ~ \textcolor{red}{\text{exp} \left( - i x S J k_{y0} \delta t \right)} \right)$
\STATE $\partial g / \partial y = F^{-1}_{k_{y} \rightarrow y} \left( F^{-1}_{k_{x} \rightarrow x} \left( \overline{\partial g / \partial y} \right) ~ \textcolor{red}{\text{exp} \left( - i x S J k_{y0} \delta t \right)} \right)$
\STATE
\STATE $\left[ \left\langle \phi \right\rangle, g_{s} \right] = v_{E x} ~ \partial g / \partial x - v_{E y} ~ \partial g / \partial y$
\STATE $\overline{\left[ \left\langle \phi \right\rangle, g_{s} \right]} = F_{x \rightarrow k_{x}} \left( F_{y \rightarrow k_{y}} \left( \left[ \left\langle \phi \right\rangle, g_{s} \right] \right) ~ \textcolor{red}{\text{exp} \left( i x S k_{y} \delta t \right)} \right)$

\end{algorithmic}
\end{algorithm}
\end{figure}

In GENE, linearly interpolating the values of $\vec{k}_{\perp}$ appearing in equations (\ref{eq:FourierExBvelocity}) and (\ref{eq:FourierDistFnGradient}) and including the phase factor in the Fourier and inverse Fourier transforms was all that was needed to eliminate the nonphysical coupling. The scalar value of $k_{\perp}$ appearing the Bessel function of equation (\ref{eq:FourierExBvelocity}) does {\it not} need to be corrected because it only affects the coupling strength and not which modes couple. Thus, like the linear terms, it is a smooth function of $k_{x}$ and any error will converge to zero with increasing $k_{x}$ resolution.

The Bessel function could be corrected, but at some computational cost. This is because the values of the Bessel function on the full five-dimensional grid are typically calculated at the start of the simulation. However, we now require their values at many locations between $k_{x}$ grid points. Exactly correcting this requires the Bessel functions to be computed at each timestep. Alternatively, approximate values could be found by interpolating between the grid points. Correcting the Bessel function will only improve the accuracy of the overall algorithm if all the linear terms and the field equations (e.g. quasineutrality) are corrected as well.


\section{Nonlinear coupling test case}

To verify the above changes to the flow shear implementation in GENE, we compared simulations against analytic results for a simple nonlinear test case. This case was designed to clearly expose any errors with the implementation of the nonlinear term in the presence of perpendicular flow shear. The analytic derivation begins with the electrostatic, collisionless gyrokinetic model in slab geometry without magnetic shear. The electrons are assumed to respond adiabatically to the motion of a single ion species. We neglect background flow and shear in the flow parallel to the magnetic field, but allow shear in the flow perpendicular to the field flow (i.e. $E \times B$ flow shear). The gradients of density and temperature are set to zero. Finally, the distribution function and electrostatic potential are initialized to be constant in the parallel direction, which ensures that the model is entirely free of variation in the parallel direction (i.e. $k_{||} = 0$).

With these simplifications, the Fourier space gyrokinetic equation (using a {\it conventional} Fourier transform) is given by \cite{ParraUpDownSym2011}
\begin{align}
  \left( \frac{\partial}{\partial t} - S k_{y} \frac{\partial}{\partial k_{x}} \right)& \overline{g}_{i}(\vec{k}) \nonumber \\
  =& \int d \vec{k}' \int d \vec{k}'' ~ \delta_{D} \left( \vec{k} - \vec{k}' - \vec{k}'' \right) \left( \vec{k}'' \times \vec{k}' \right) \cdot \frac{\hat{b}}{B} \label{eq:GKeq} \\
  &\times \left( \overline{g}_{i}(\vec{k}') + \frac{Z_{i} e F_{Mi}}{T_{i}} \overline{\phi}(\vec{k}') J_{0}\left(k' \rho_{i}\right) \right) \overline{\phi}(\vec{k}'') J_{0}\left(k'' \rho_{i}\right) , \nonumber
\end{align}
where the integrals are performed over the full $k_{x} \in \left( - \infty, \infty \right)$, $k_{y} \in \left( - \infty, \infty \right)$ plane and $\delta_{D}$ is the Dirac delta function. Here $Z_{i}$ is the ion charge number, $e$ is the elementary charge, $T_{i}$ is the ion temperature, and $F_{Mi}$ is the ion Maxwellian distribution function. This is simply the Fourier space version of equation (\ref{eq:gksymPerturbed}), where the linear and drive terms have all vanished due to our assumptions. The gyrokinetic model is closed by the quasineutrality equation,
\begin{align}
  \int d^{3}v ~ \overline{g}_{i}(\vec{k}) J_{0}\left(k \rho_{i}\right) = \frac{e n_{i}}{T_{e}} C_{Q}(k) \overline{\phi}(\vec{k}) , \label{eq:QNeq}
\end{align}
where $n_{i}$ is the ion particle density, $T_{e}$ is the electron temperature,
\begin{align}
  C_{Q}(k) \equiv Z_{i} \frac{T_{e}}{T_{i}} \left[ 1 - I_{0} \left( k^{2} \rho_{th i}^{2} \right) \exp \left( - k^{2} \rho_{th i}^{2} \right) \right]+ 1 ,
\end{align}
$\rho_{th i} \equiv \sqrt{T_{i} / m_{i}} / \Omega_{i}$ is the ion thermal gyroradius, $m_{i}$ is the ion mass, and $\Omega_{i}$ is the ion gyrofrequency.

Next, we rename the dummy variables $\vec{k}' \rightarrow \vec{k}''$ and $\vec{k}'' \rightarrow \vec{k}'$ in equation (\ref{eq:GKeq}) and sum the resulting equation with the original equation (\ref{eq:GKeq}). Then, we will introduce the coordinate system transform $\vec{K} \equiv \vec{k} + S k_{y} t \hat{e}_{x}$ and first note that $\vec{k}'' \times \vec{k}' = \vec{K}'' \times \vec{K}'$. Additionally, we will restrict ourselves to a discrete grid of $\vec{K}$ values as is necessary in numerical simulations. This gives
\begin{align}
  \left. \frac{\partial}{\partial t} \right|_{K} & \overline{g}_{i}(\vec{K}) \label{eq:GKeqMod} \\
  &= \frac{1}{2} \sum_{\vec{K}'} \left( \vec{K}'' \times \vec{K}' \right) \cdot \frac{\hat{b}}{B} \left( \overline{g}_{i}(\vec{K}') \overline{\phi}(\vec{K}'') J_{0}\left(k'' \rho_{i}\right) - \overline{g}_{i}(\vec{K}'') \overline{\phi}(\vec{K}') J_{0}\left(k' \rho_{i}\right) \right) , \nonumber
\end{align}
where the summation is still over the full grid and the Dirac delta function has been replaced by the wavevector coupling condition of $\vec{K}'' = \vec{K} - \vec{K}'$. Note that, since the time derivative is now taken at constant $\vec{K}$, the quantities of $k \rho_{i}$ that appear as arguments to the Bessel function have acquired a time dependence.

Now, as in a three-wave resonant decay calculation \cite{HasegawaThreeWaveCoupling1979}, we will consider just three Fourier modes with $\vec{K}_{1} = \vec{K}_{2} + \vec{K}_{3}$ such that they nonlinearly couple. We will initialize $\overline{\phi}_{2} \sim \overline{\phi}_{3} \gg \overline{\phi}_{1}$ (where the subscript specifies the Fourier mode) and linearize equation (\ref{eq:GKeqMod}) in the ratio of the mode amplitudes. We find
\begin{align}
  \left. \frac{\partial}{\partial t} \right|_{K} \overline{g}_{i2} &= 0 \\ 
  \left. \frac{\partial}{\partial t} \right|_{K} \overline{g}_{i3} &= 0
\end{align}
for the two ``pump'' modes, indicating that both distribution functions remain constant in time. Thus, if both distribution functions are initialized to be Maxwellian according to $\overline{g}_{i} \propto n_{i} (m_{i}/(2 \pi T_{i}))^{3/2} \exp (-m_{i} v^{2} / (2 T_{i}))$, then they will remain so. Then, we can substitute a Maxwellian into equation (\ref{eq:QNeq}), use the identity
\begin{align}
    \int_{0}^{\infty} dv_{\perp} v_{\perp} \exp ( - a^{2} v_{\perp}^{2}) J_{0} (k v_{\perp}) = \frac{1}{2 a^{2}} \exp \left( -\left( \frac{k}{2 a} \right)^{2} \right), 
\end{align}
and calculate the corresponding potentials to be
\begin{align}
  \overline{\phi}_{2} &= \hat{\phi}_{2}^{\varnothing} C_{Q2}^{-1} \exp \left( - \frac{1}{2} k_{2}^{2} \rho_{thi}^{2} \right) \label{eq:mode2sol} \\ 
  \overline{\phi}_{3} &= \hat{\phi}_{3}^{\varnothing} C_{Q3}^{-1} \exp \left( - \frac{1}{2} k_{3}^{2} \rho_{thi}^{2} \right) \label{eq:mode3sol} ,
\end{align}
where $\hat{\phi}_{j}^{\varnothing} \equiv \overline{\phi}_{j}^{\varnothing} C_{Qj}^{\varnothing} \exp \left( \frac{1}{2} k_{j}^{\varnothing 2} \rho_{thi}^{2} \right)$, the superscript ``$\varnothing$'' indicates the initial value, and we note that $\vec{k}_{j}^{\varnothing} = \vec{K}_{j}$.

The evolution of mode 1, the ``driven'' mode, is more complex. The linearized gyrokinetic equation for it is
\begin{align}
  \left. \frac{\partial}{\partial t} \right|_{K} \overline{g}_{i1} &= \left( \vec{K}_{3} \times \vec{K}_{2} \right) \cdot \frac{\hat{b}}{B} \left[ \overline{g}_{i2} \overline{\phi}_{3} J_{0}\left(k_{3} \rho_{i}\right) - \overline{g}_{i3} \overline{\phi}_{2} J_{0}\left(k_{2} \rho_{i}\right) \right] \label{eq:endDiffEq} .
\end{align}
Since the solutions of modes 2 and 3 are given by equations (\ref{eq:mode2sol}) and (\ref{eq:mode3sol}), we can substitute them and directly integrate to find $\overline{g}_{i1}$. This can then be substituted into the quasineutrality equation (i.e. equation (\ref{eq:QNeq})) to find the solution
\begin{align}
  \overline{\phi}_{1} = \frac{1}{n_{i}} \left( \vec{K}_{3} \times \vec{K}_{2} \right) & \cdot \frac{\hat{b}}{B} \frac{\hat{\phi}_{2}^{\varnothing} \hat{\phi}_{3}^{\varnothing}}{C_{Q1}} \int d^{3}v \Bigg\{ F_{Mi} J_{0}\left(k_{1} \rho_{i} \right) \nonumber \\
  \times \left. \int_{0}^{t} \right|_{K} d t' &\left[ C_{Q3}'^{-1} \exp \left( - \frac{1}{2} k_{3}'^{2} \rho_{thi}^{2} \right) J_{0}\left(k'_{3} \rho_{i}\right) \right. \\
  &- \left. C_{Q2}'^{-1} \exp \left( - \frac{1}{2} k_{2}'^{2} \rho_{thi}^{2} \right) J_{0}\left(k'_{2} \rho_{i}\right) \right] \Bigg\} , \nonumber
\end{align}
where the prime indicates that the quantity is evaluated at $t'$, not $t$. Like the derivative in equation (\ref{eq:endDiffEq}), the $|_{K}$ in the integral indicates that $\vec{K}$ is held fixed, rather than $\vec{k}$. This equation is simple to solve numerically. However, by choosing wavenumbers such that $k \rho_{i} \ll 1$, we can expand to lowest order and make use of $J_{0} (x) \approx 1 - x^{2}/4$, $\exp (x) \approx 1 + x$, and $I_{0} (x) \exp (-x) \approx 1 - x$. This produces the much simpler result of
\begin{align}
  \overline{\phi}_{1} = \left( 1 + \frac{T_{i}}{Z_{i} T_{e}} \right) \left( \vec{K}_{3} \times \vec{K}_{2} \right) & \cdot \frac{\hat{b}}{B} \overline{\phi}_{2}^{\varnothing} \overline{\phi}_{3}^{\varnothing} \left( 1 + k_{1}^{2} \rho_{S}^{2} \right)^{-1} \left. \int_{0}^{t} \right|_{K} d t' \left( k_{2}'^{2} - k_{3}'^{2} \right) \rho_{S}^{2} , \label{eq:finalSol}
\end{align}
where $\rho_{S} \equiv (1/\Omega_{i}) \sqrt{Z_{i} T_{e}/m_{i}}$ is the sound gyroradius and the integral can be found to be
\begin{align}
  \left. \int_{0}^{t} \right|_{K} d t' \left( k_{2}'^{2} - k_{3}'^{2} \right) = \left( \frac{1}{3} \left( K_{2y}^{2} - K_{3y}^{2} \right) S^{2} t^{2} - \left( K_{2x} K_{2y} - K_{3x} K_{3y} \right) S t + K_{2}^{2} - K_{3}^{2} \right) t .
\end{align}
This solution is in agreement with reference \cite{HasegawaThreeWaveCoupling1979}, if the cold ion limit (i.e. $T_{i}/(Z_{i} T_{e}) \ll 1$) is taken and the flow shear is set to zero.

\begin{figure}
  \centering
  \includegraphics[width=\textwidth]{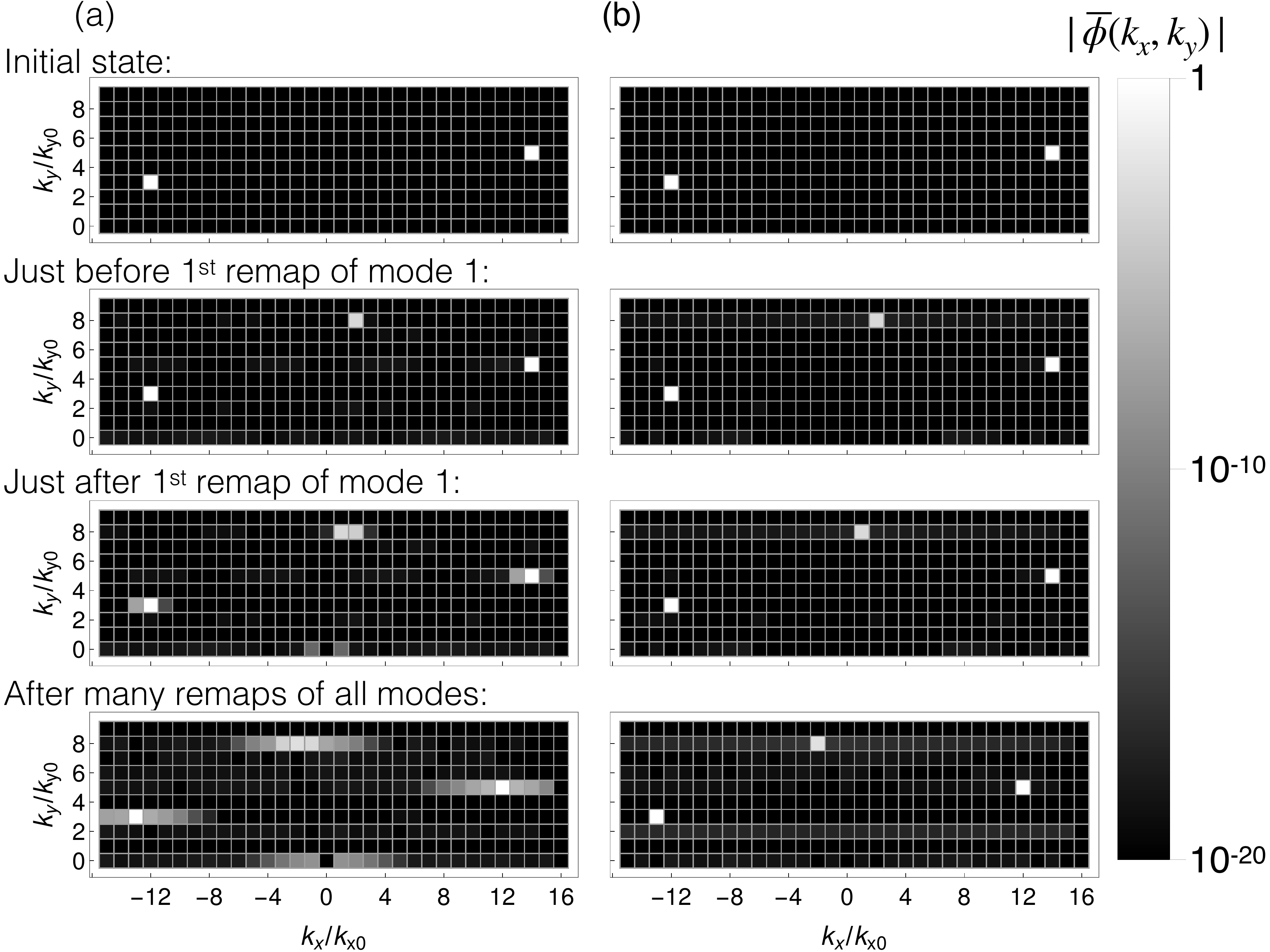}
  \caption{Spectra of $\left| \overline{\phi} \right|$ calculated by the (a) original wavevector-remap method and the (b) corrected wavevector-remap method for the simple test case of two large pump modes nonlinearly coupling to drive the growth of a third mode.}
  \label{fig:spectra}
\end{figure}

This result can be compared to a GENE calculation, given the same assumptions. This is fairly straightforward to do, with the one caveat that the flux surface averaged value of the electrostatic potential must be set to zero when the code calculates the adiabatic electron response. This is because there are no flux surfaces in slab geometry when the magnetic shear is zero -- every field line closes on itself at the ends of the domain. With this term neglected, we can initialize two large pump modes as shown in the top row of figure \ref{fig:spectra}. For this specific simulation $k_{x0} \rho_{thi} = 5 \times 10^{-3}$, $k_{y0} \rho_{thi} = 0.01$, $Z_{i} = 1$, $T_{i} = T_{e}$, $\vec{K}_{2} = (-12 k_{x0}, 3 k_{y0})$, $\vec{K}_{3} = (14 k_{x0}, 5 k_{y0})$, $\vec{K}_{1} = \vec{K}_{2} + \vec{K}_{3} = (2 k_{x0}, 8 k_{y0})$, and $S = 1.6 \times 10^{-3} L_{||}/v_{thi}$ (where $L_{||}$ is the length of the box in the parallel direction). Subsequent rows of figure \ref{fig:spectra} demonstrate that unphysical nonlinear coupling occurs in the original wavevector-remap scheme and causes a smearing of the mode amplitudes in Fourier space. This does not occur in the corrected wavevector-remap method. Instead, the two pump modes couple together to drive the same single mode at all times.

\begin{figure}
  \centering
  \includegraphics[width=\textwidth]{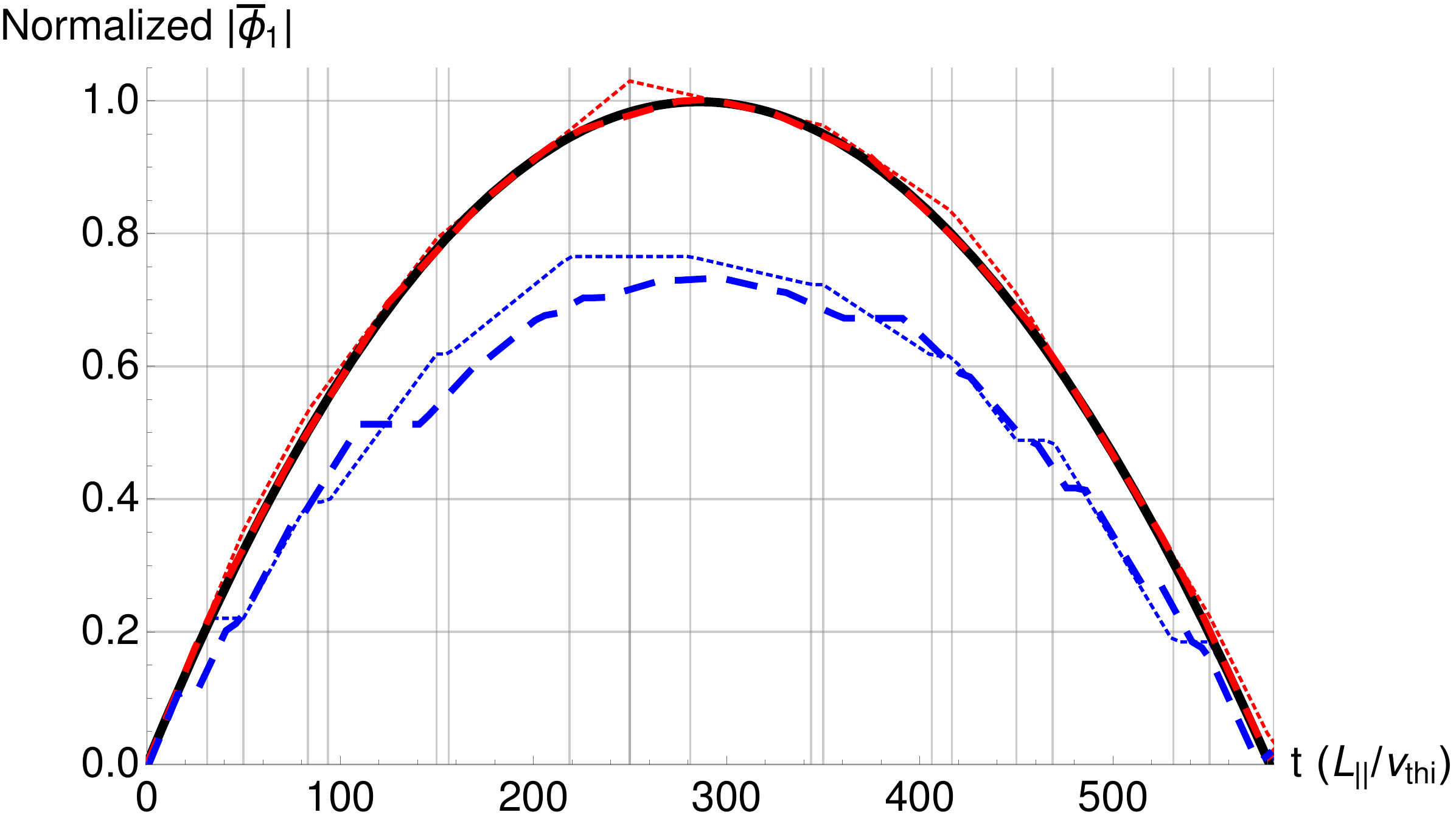}
  \caption{The amplitude of the driven mode calculated by equation (\ref{eq:finalSol}) (black), the original wavevector-remap method (blue), and the corrected wavevector-remap method (red) for the test case shown in figure \ref{fig:spectra}. Note the thin dotted lines use the wavenumber resolution shown in figure \ref{fig:spectra}, while the thick dashed lines have twice the $k_{x}$ resolution. The vertical grid lines indicate when any of the three modes are remapped in the low $k_{x}$ resolution simulation (i.e. the thin dotted lines).}
  \label{fig:amplitude}
\end{figure}

Figure \ref{fig:amplitude} shows the amplitude of the driven mode (i.e. $\vec{k}_{1}$) with time. We see that, even at low $k_{x}$ resolution, the corrected wavevector-remap method well approximates the analytical solution. There is a small error, which arises from the Bessel functions (as detailed in the last two paragraphs of section III), but even this vanishes at higher $k_{x}$ resolution. This verifies our implementation. In contrast, the original wavevector-remap method converges to an incorrect solution and, looking closely, it is apparent why. Because the remaps are occurring at different times for the different modes, the pump modes are only coupled to the driven mode part of the time. We see that, at the start of the simulation, the mode has the proper evolution. However, after the driven mode is remapped, the pump modes remain coupled to its original location, so it stagnates. Then, after one of the pump modes is remapped, it resumes its growth. This process repeats again and again as the simulation progresses.

\section{Effect of the remap scheme in standard gyrokinetic test cases.}

We perform a comparison of the original and corrected remap methods using two conventional tokamak parameter sets, widely used for benchmarking purposes. These are, firstly, the  {\it CYCLONE} case \cite{DimitsCycloneBaseCase2000} and secondly the {\it Waltz standard case}\cite{Waltzcase} (parameters are described below). The most striking difference between the CYCLONE and Waltz case is possibly that the linear eigenmodes are much more elongated along the field line in the Waltz case, so that the CYCLONE case tends to produce radially elongated streamer-like structures on the outboard mid-plane during the linear phase of simulations, but the Waltz case produces modes with large radial modulation. The streamers saturate when they are broken up by zonal flows in ITG simulations, whereas the nonlinear saturation of modes in the Waltz case is only partly due to zonal flows. The difference in nonlinear saturation physics between CYCLONE and Waltz cases allows testing of the nonlinear physics effects of remap algorithms in two quite different regimes.

In both of these test cases a background toroidal rotation shear is applied; the size of the resulting $E \times B$ shear $S$ in these simulations is $0.12 v_t/R$, where $v_t$ is the thermal velocity of the kinetic species ($v_t= \sqrt{T/m}$) and $R$ is the device major radius. These cases are run with original remap and corrected remap, and, for the CYCLONE case, in both the ITG (electrons treated as adiabatic) and ETG (ions treated as adiabatic) model; for the Waltz case we only perform ETG simulations. In the ETG case, the ion adiabatic response is a uniform Boltzmann response, whereas in the ITG case the electron response is only Boltzmann on each magnetic surface, so the electrons do not respond adiabatically to zonal potential . The result is that the ETG simulations have a weaker zonal flow drive\citep{JenkoGENE2000}. As part of this exercise, a simulation box-size scan (scaling system size uniformly in the radial and binormal directions) is performed for the Waltz case. 

Performing with ITG and ETG simulations allow us to emphasise two extreme limits: the ITG case where dominant nonlinear interactions are between the zonal modes and the finite $k_{y}$ drift modes and the ETG case where direct interactions between various finite $k_{y}$ modes are important. We therefore expect that the issues related to the original remap method will affect the ITG case less than the ETG case, since the off-by-one errors in the nonlinear coupling do not impact on interactions involving the zonal modes. The shearing rate chosen for this numerical study is larger than typical physical values for the ETG case.

Several diagnostics are selected for discussion, including the heat fluxes and zonal electrostatic potential.
An additional diagnostic that probes turbulence statistical properties is the kurtosis of the binormal electric field component $E_y = -\partial \phi / \partial y$ on the outboard mid-plane. For a quantity $X$ with zero mean, the kurtosis is defined as $ \langle X^4 \rangle /  \langle X^2 \rangle^2$ (with the angle brackets denoting a mean which is here both a spatial average over the $y$ direction and a time average over the last half of the simulation), and in some sense measures the heaviness of the tails of the distribution of $X$. The kurtosis has a value of $3$ for Gaussian distributed data, and higher if long tails are present. The idea is that the true local turbulence intensity (here the binormal electric field is a proxy for this) is expected to be somewhat intermittent, and depart from the Gaussian distribution expected if all the waves added up with random independent phases. The correlations between wave phases are thus in some sense associated with `long tails' in the histogram of local quantities. Thus, the kurtosis should be sensitive to the tendency of the original remap scheme to unphysically remove phase correlations.

\subsection{CYCLONE case}

\begin{figure}[htb]
\centering
\epsfig{figure=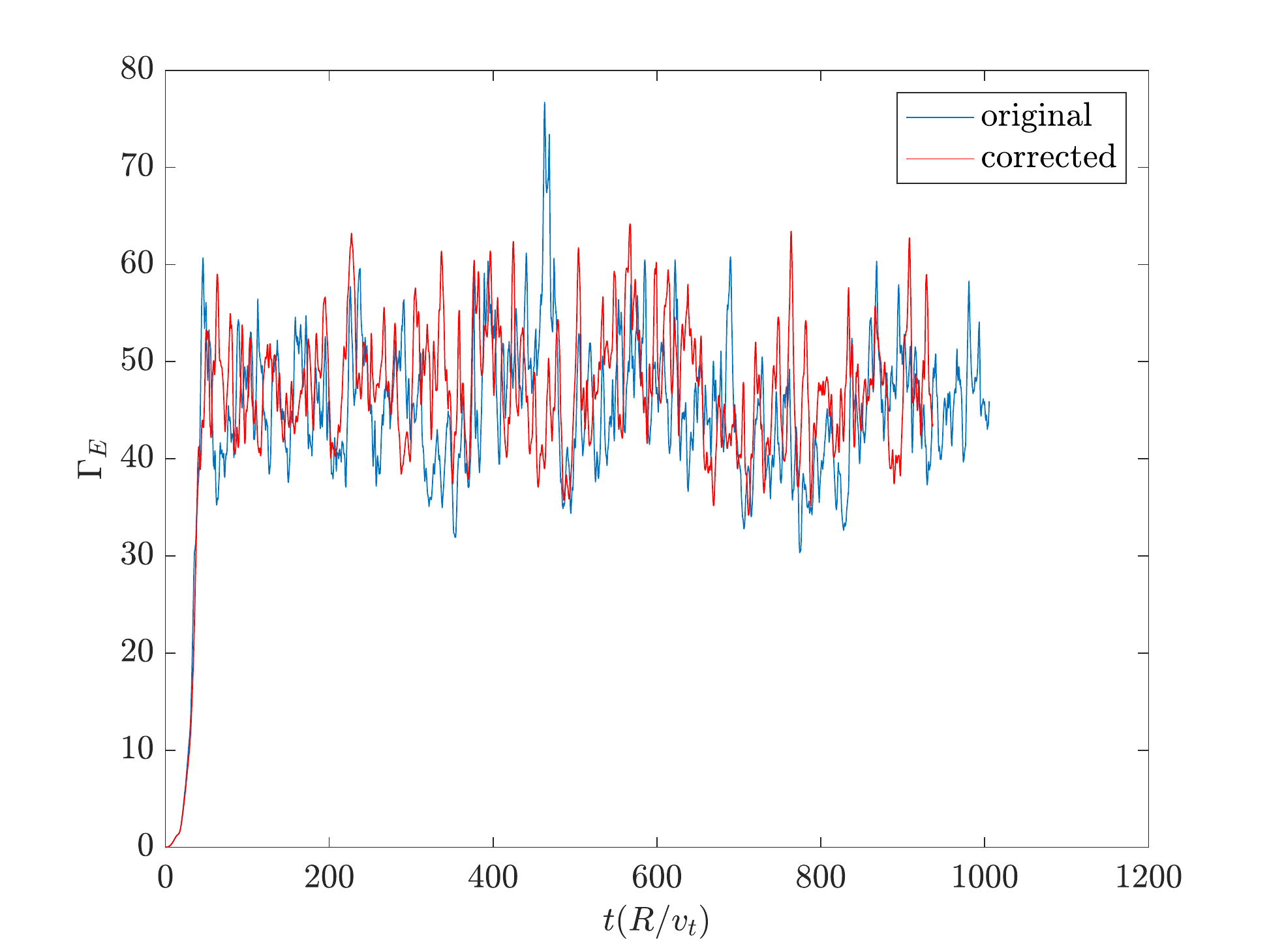,width=11cm}
\caption{Heat flux (in units of $n m v_t^3 \rho_t/R$) versus time in CYCLONE ITG case.}
  \label{fig:heatflux_cyclone}
\end{figure}

The CYCLONE case is designed to represent, in simplified geometry, the core parameters of a DIII-D discharge, and has been widely used as a reference case to test gyrokinetic codes and examine basic core turbulence physics. The simulations use equal ion and electron temperatures, safety factor $q=1.4$, shear $(dq/dr) r/q = 0.8$, density and temperature gradient length scales $L_n = R/2.2$ and $L_T = R/6.0$ respectively, and a concentric circular equilibrium, with local aspect ratio $0.18$. 

This simulation uses a box size of $L_x=175 \rho_{t}$ and $L_y = 125 \rho_{t}$ (where $\rho_t = v_t/\Omega$ is the thermal gyroradius for a particle with gyrofrequency $\Omega$). The maximum wavenumbers resolved are $k_x \rho_t=3.2$ and $k_y \rho_t= 2.0$. These numerical parameters are typical of more well-resolved CYCLONE simulations.

Figure \ref{fig:heatflux_cyclone} shows that the initial ITG fluxes are very high, but late time fluxes resolve a gyroBohm diffusivity of around $1$. Note that since only the nonlinear terms are modified between the original and corrected simulations, linear simulations are identical, and the initial time traces only diverge once the amplitudes become sufficiently large.

\begin{figure}
  \centering
  \includegraphics[width=\textwidth]{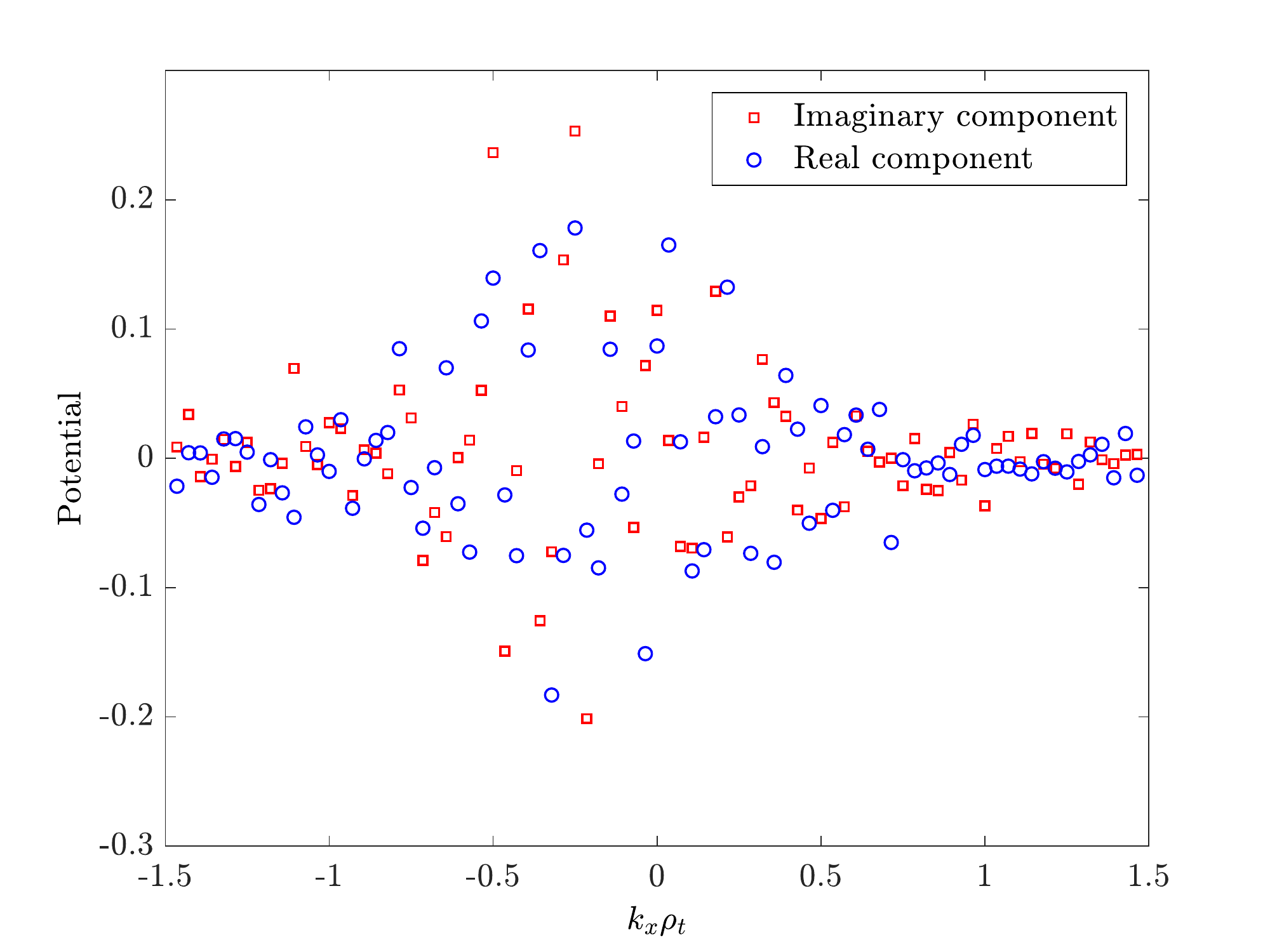}
  \caption{A snapshot of the real and imaginary Fourier coefficients of the electrostatic potential at the outboard mid-plane $k_x$ for $k_y \rho_t= 0.2$.}
  \label{fig:fouriermodes_cyclone}
\end{figure}

Figure \ref{fig:fouriermodes_cyclone} shows the Fourier components of the potential versus $k_x$ for $k_y\rho_{t} =0.2$ at the outboard mid-plane at the end of this simulation. The envelope of these points peaks at negative $k_x$ due to the homogeneous flow shear. Note that the points, representing the real and imaginary components of the potential are scattered within a rough envelope, but there is no obvious correlation between neighbouring points. This demonstrates the discontinuous nature of the Fourier components versus grid point index, unlike the spatial functions, which are smooth on scales finer than the dissipation scale, where the (numerical or collisional) phase-space diffusivity dominates the dynamics, suppressing variations in the distribution function. The lack of phase correlations between neighbouring Fourier modes is a consequence of the spatial translation invariance of the gyrokinetic theory (this is actually a discrete symmetry once parallel boundary conditions are taken into account): since spatial translation in the radial direction rotates the relative phase between modes with different $k_x$, no particular relative phase is preferred. Even the Fourier transform of a Gaussian peak, localised at $x = L_x/2$, would have mode phases that oscillate by a factor of $\pi$, so the non-smoothness of Fourier coefficients is not really a specific consequence of turbulence.

\begin{figure}
\begin{subfigure}{0.48\textwidth}
\includegraphics[width=8.5cm]{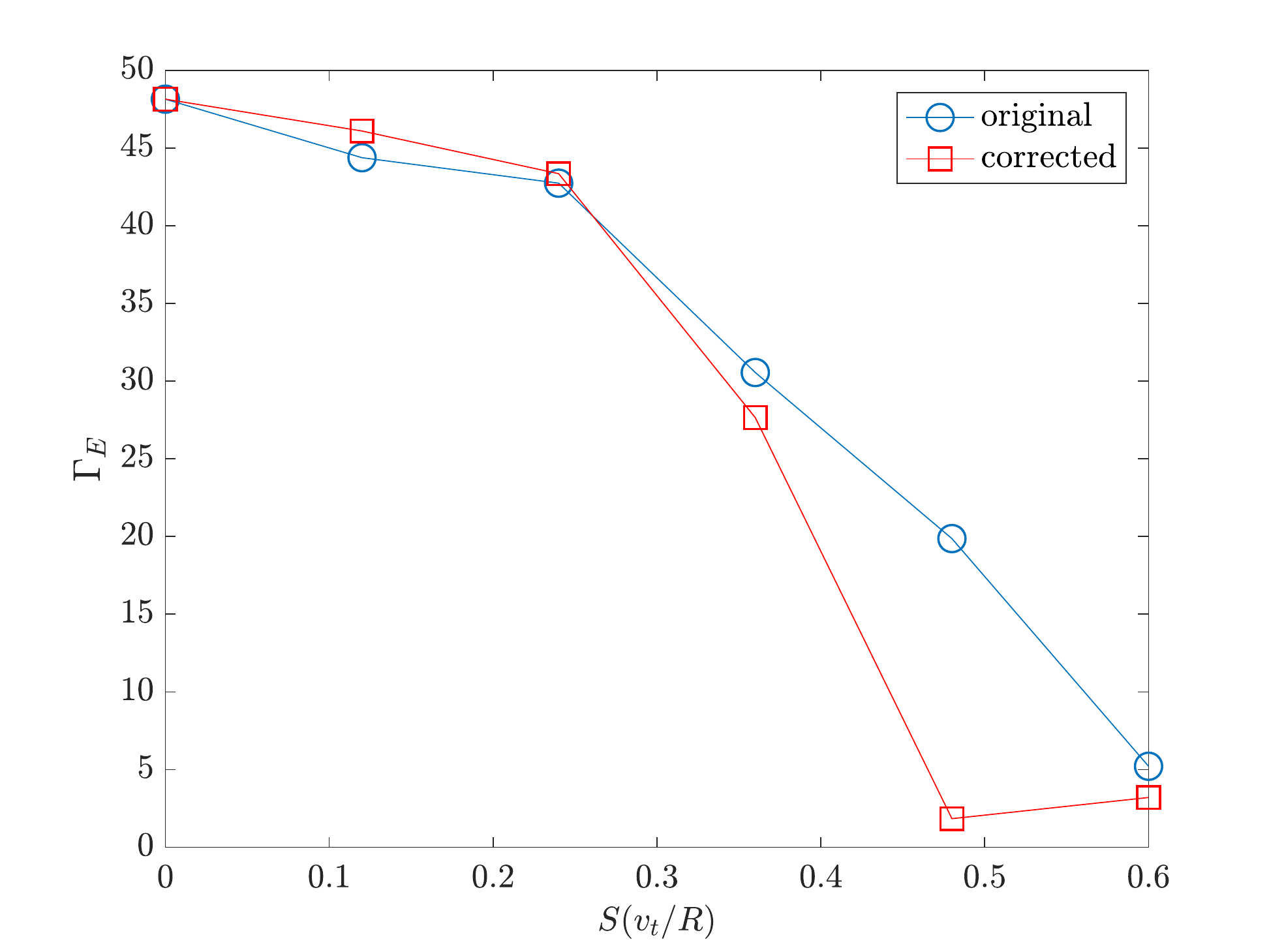} \caption{} \label{fig:cyclone_fluxscan}
\end{subfigure}
\begin{subfigure}{0.48\textwidth}
\includegraphics[width=8.5cm]{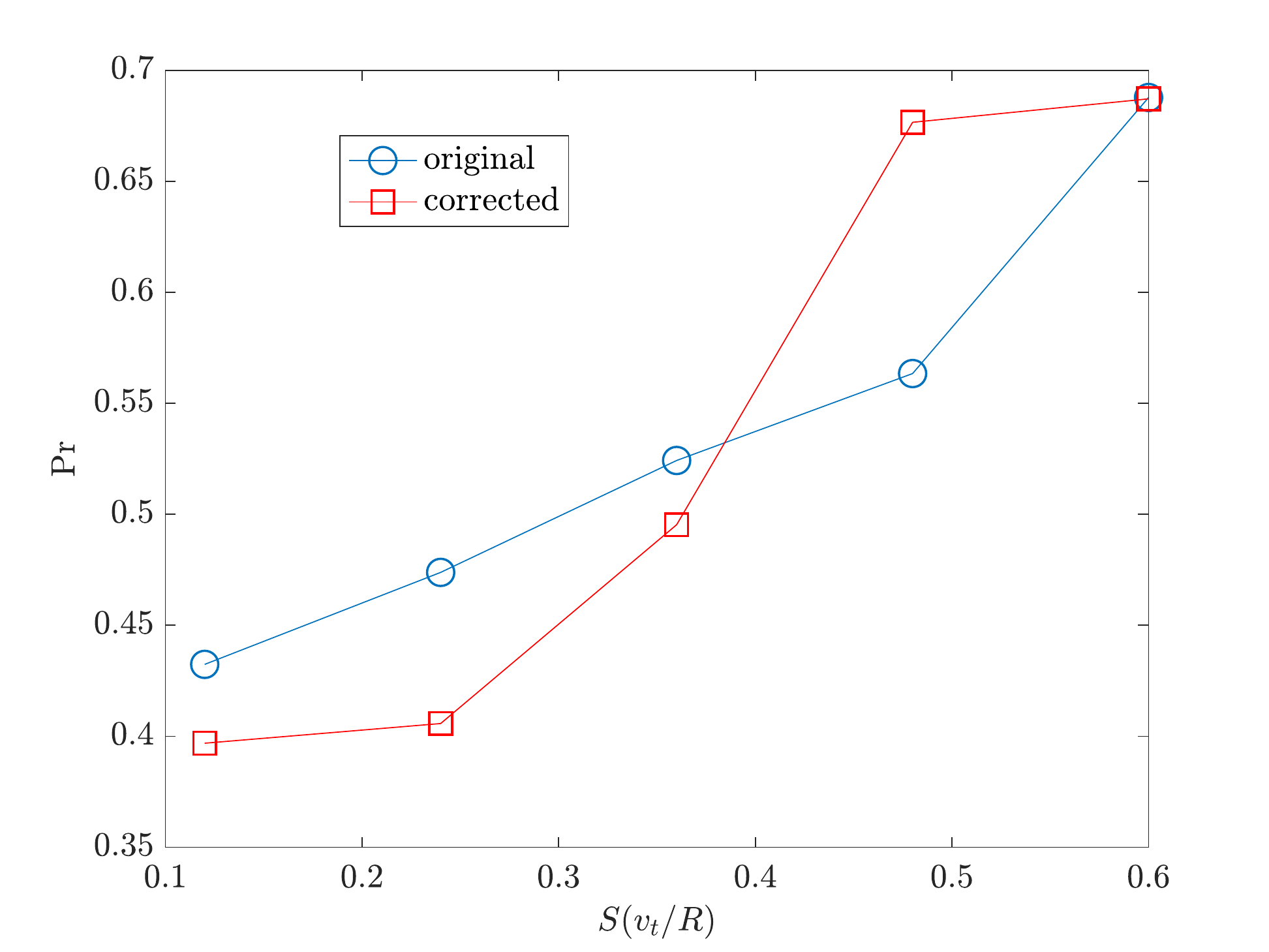} \caption{} \label{fig:cyclone_prandtl}
\end{subfigure}

\caption{(a) Heat flux versus shearing rate $S$ and (b) Prandtl number $\operatorname{Pr}=\chi_v/\chi_E$ for CYCLONE ITG simulations.
}
\label{fig:scan_cyc}
\end{figure}

A primary objective of tokamak gyrokinetic simulation is to predict fluxes due to turbulence, and remap simulations have principally been used to understand how heat and momentum fluxes are modified by sheared rotation\cite{Casson,kinsey}. We performed a scan in shearing rate over $S=[0,0.6] v_t/R$, with parameters otherwise unmodified, for the CYCLONE ITG case. For the CYCLONE case, increasing shearing rate leads to almost complete turbulence quenching\cite{Roach_GF_for_ST}, despite the onset of the $v_{||}$ instability\cite{Catto_Rosenbluth} at large shearing rates. We find that levels of flux are generally comparable between the original and corrected wavevector remap simulations; the transition to near-zero flux due to background shear stabilisation occurs at somewhat lower shearing rate in the corrected-remap simulations (this shift is also consistent with simulations using other non-remap methods\cite{candy_remap}). Note that other choices related to the numerical method, such as using periodic versus absorbing boundary conditions\citep{mcmillan_pringle_teaca_2018}, can also make a substantial difference to these curves; this sensitivity is at least somewhat connected to the question of subcriticality.   

\begin{figure}[htb]
\centering
\begin{subfigure}{0.48\textwidth}
\includegraphics[width=8.5cm]{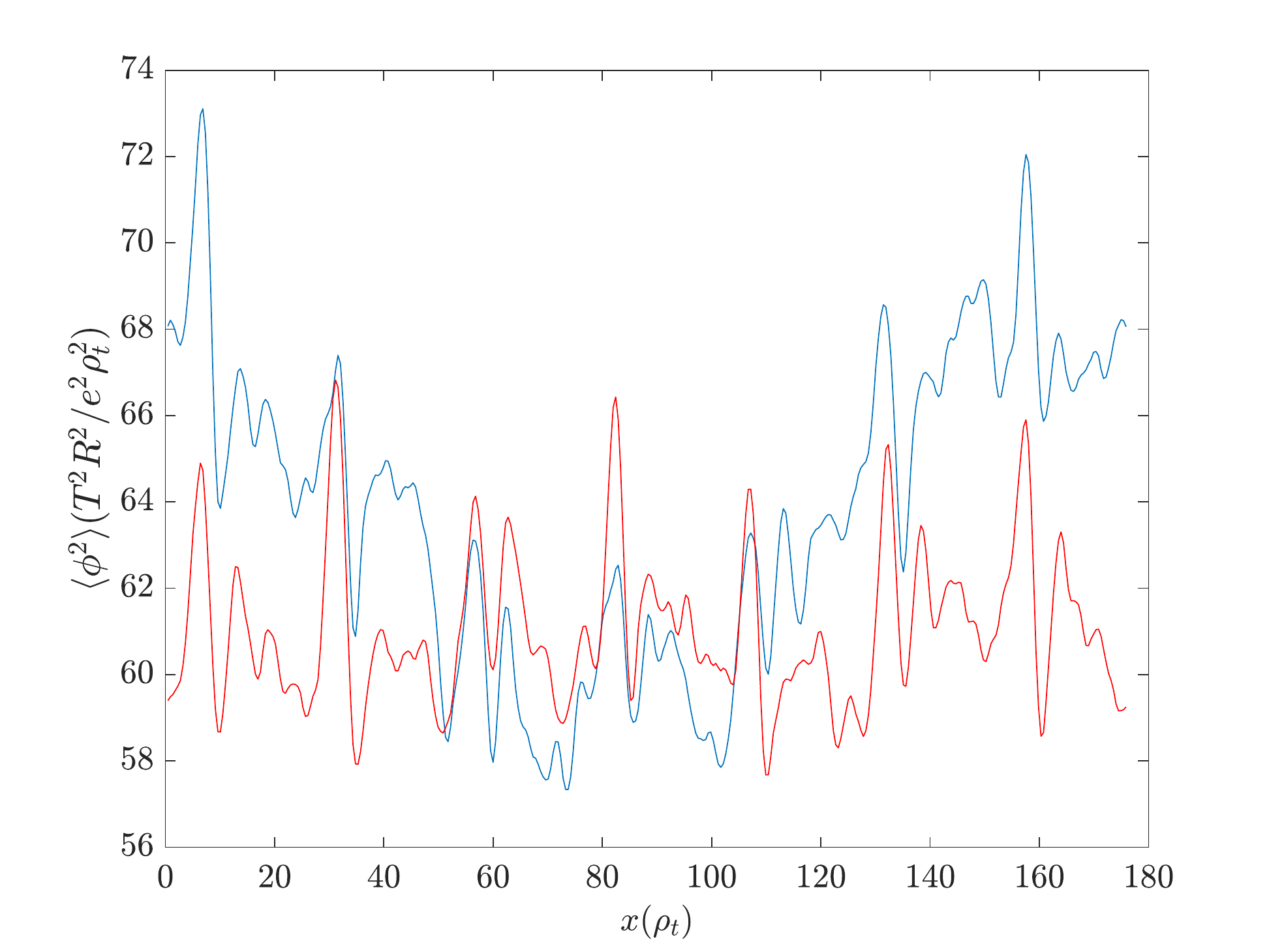} \caption{} \label{fig:intensvary}
\end{subfigure}
\begin{subfigure}{0.48\textwidth}
\includegraphics[width=8.5cm]{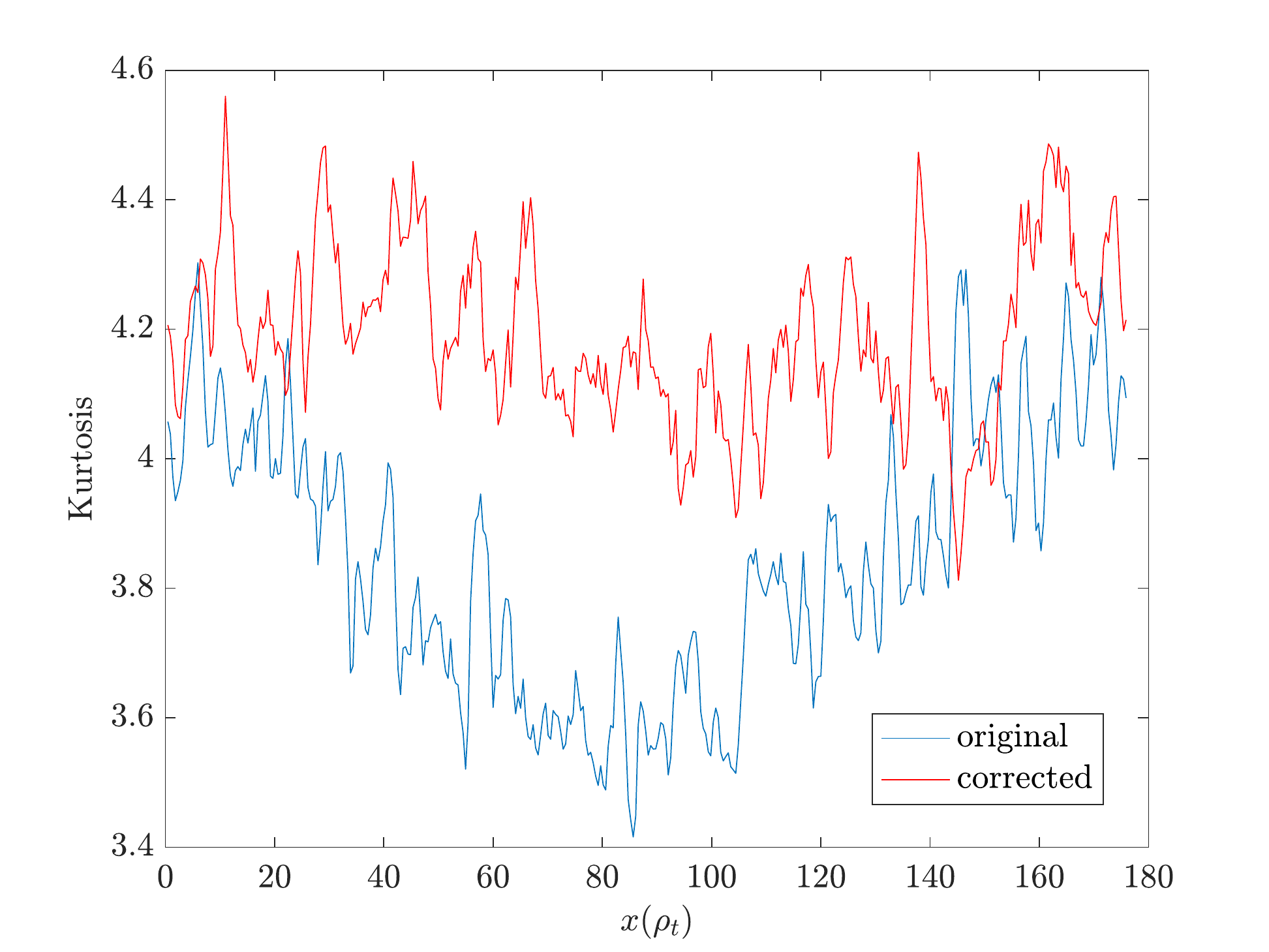} \caption{} \label{fig:kurtosis_cyc036}
\end{subfigure}
\caption{(a) Turbulence intensity (mean squared value of the non-zonal potential on the mid-plane) and (b) Kurtosis versus radius for the CYCLONE ITG case with $S=0.36 v_t/R$, for the original-remap (blue) and corrected-remap (red) simulations.}
  \label{fig:radialvary_cyc}
\end{figure}

We use the Kurtosis and a measure of turbulence intensity to determine how the properties of the turbulence vary in the radial ($x$) direction. In order to retain periodicity in the $x$ and $y$ directions as well as the connection along the $z$ direction, which involves a sheared connection along the field line, it is necessary to set $L_x = L_y N / 2 \pi \hat{s}$\cite{BeerBallooningCoordinates1995}, where $N$ is an integer that specifies the number of lowest-order magnetic rational surfaces (where the magnetic field connects to itself after one turn in the poloidal direction) that fit in the radial domain. The local gyrokinetic equations have a discrete translation symmetry $x \rightarrow x + L_x/2 \pi \hat{s}$. For the CYCLONE simulations, we have $N=7$. We expect the statistical properties of the turbulence to be consistent with this symmetry, so we should not see a systematic generation of a mode with wavelength equal to the simulation box.

The turbulence intensity and Kurtosis measure (fig \ref{fig:radialvary_cyc}) of the corrected-remap simulation (the CYCLONE ITG case with $S=0.36$) are relatively uniform, with peaks in the turbulence intensity which are consistent with the seven-fold-radial symmetry. The original-remap simulation, however, shows a variation of these measures on the simulation box scale. The radial variation in the corrected remap is consistent with the underlying symmetry of the equations, but the original-remap simulation is not.

\begin{figure}[htb]
\centering
\epsfig{figure=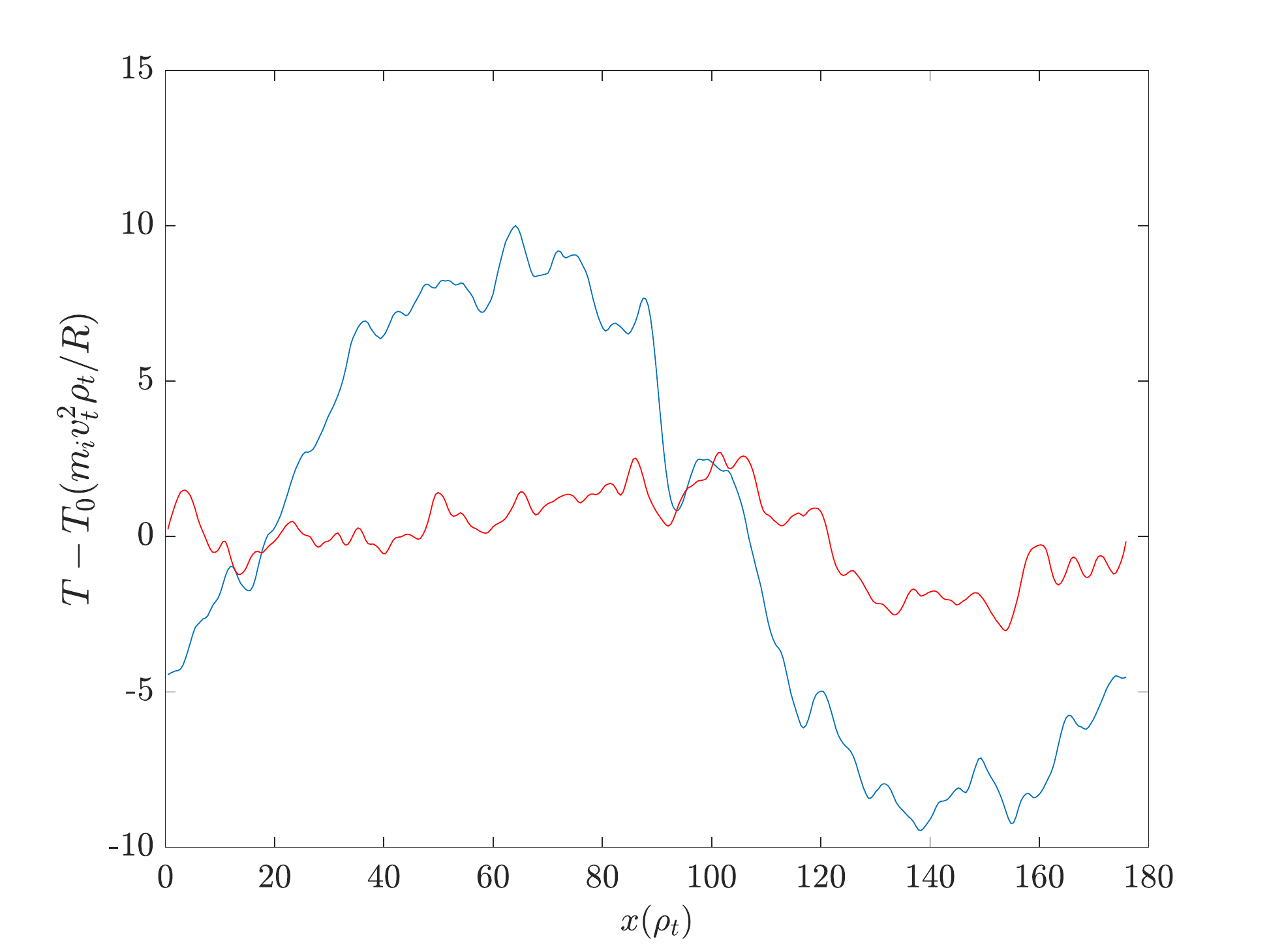,width=11cm}
\caption{Temperature fluctuation $T-T_0$ versus radius for the CYCLONE ITG case with $S=0.36 v_t/R$, for the original-remap (blue) and corrected-remap (red) simulations.}
  \label{fig:tvary}
\end{figure}

A Kurtosis measure of $3$ would be consistent with fully uncorrelated mode phases, whereas the value of around $4.2$ resolved in the corrected-remap simulation suggests that turbulence is somewhat intermittent (intensity has long tails). The effect of the off-by-one error in the original-remap is to cause jumps in the mode phase at the centre of the domain, which are likely to cause mode phases to become more uncorrelated, and this is consistent with the reduction in the Kurtosis in the centre of the $x$ domain of the original-remap simulation. The box length-scale modulation of the turbulence intensity is only around $10\%$ in the original-remap simulation, which is enough to cause much stronger structure generation than in a correct simulation. The temperature fluctuation radial profile (fig \ref{fig:tvary}) also has a strong radial variation, consistent with the variation in turbulence intensity, in order to allow for a quasi-steady state where the overall heat flux is radially constant. Based on the corrugation to the temperature profile, the box-scale variation of the temperature gradient $R/L_T$ in the original-remap simulation is $\sim 0.3$: a reduction in $R/L_T$ of $1$ would bring this simulation to the threshold for sustained turbulence, so this is a fairly significant variation in gradient.

\begin{figure}
\begin{subfigure}{0.48\textwidth}
\includegraphics[width=8.5cm]{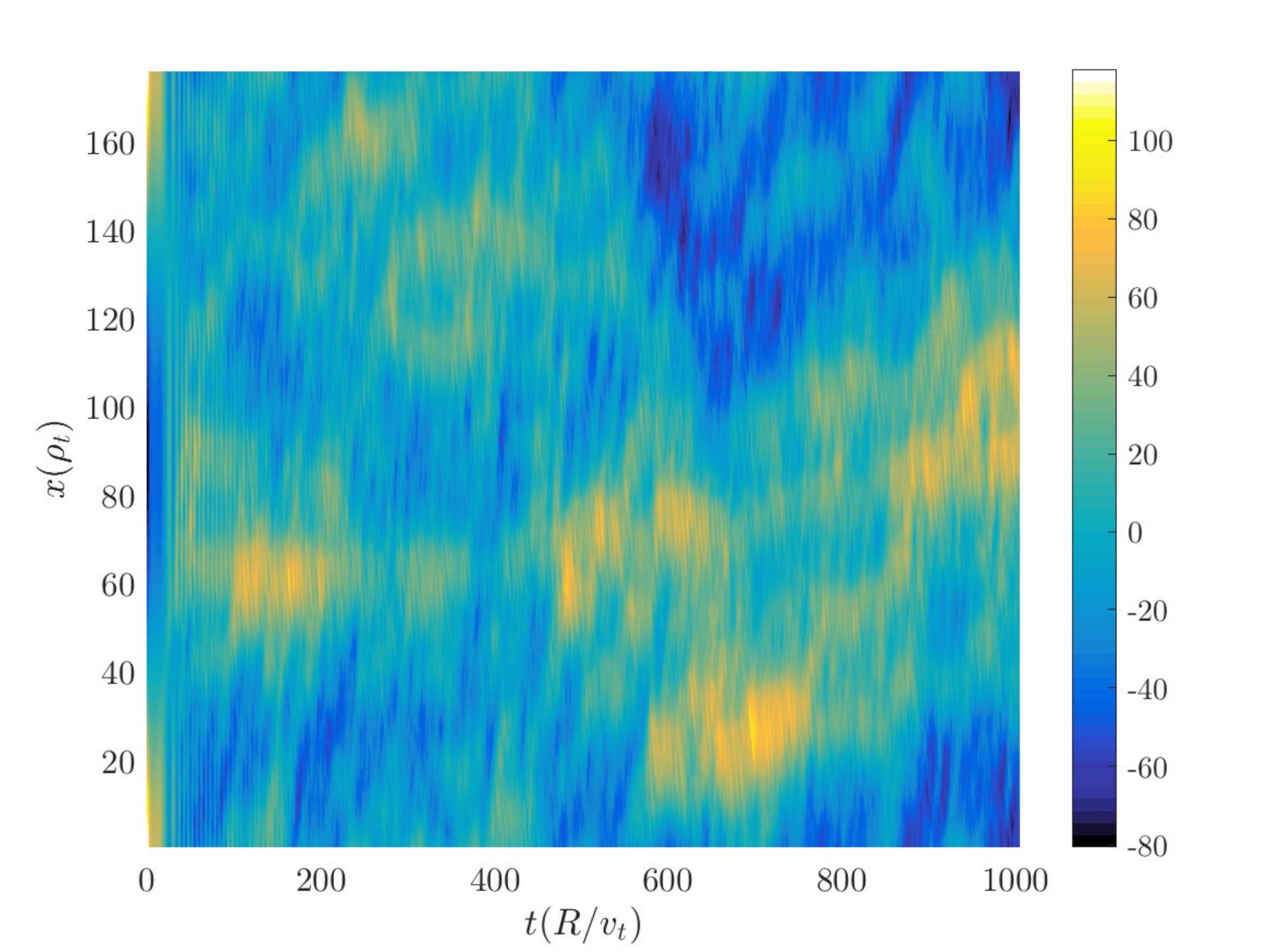} \caption{} \label{fig:zonal_original_cyc_itg2}
\end{subfigure}
\begin{subfigure}{0.48\textwidth}
\includegraphics[width=8.5cm]{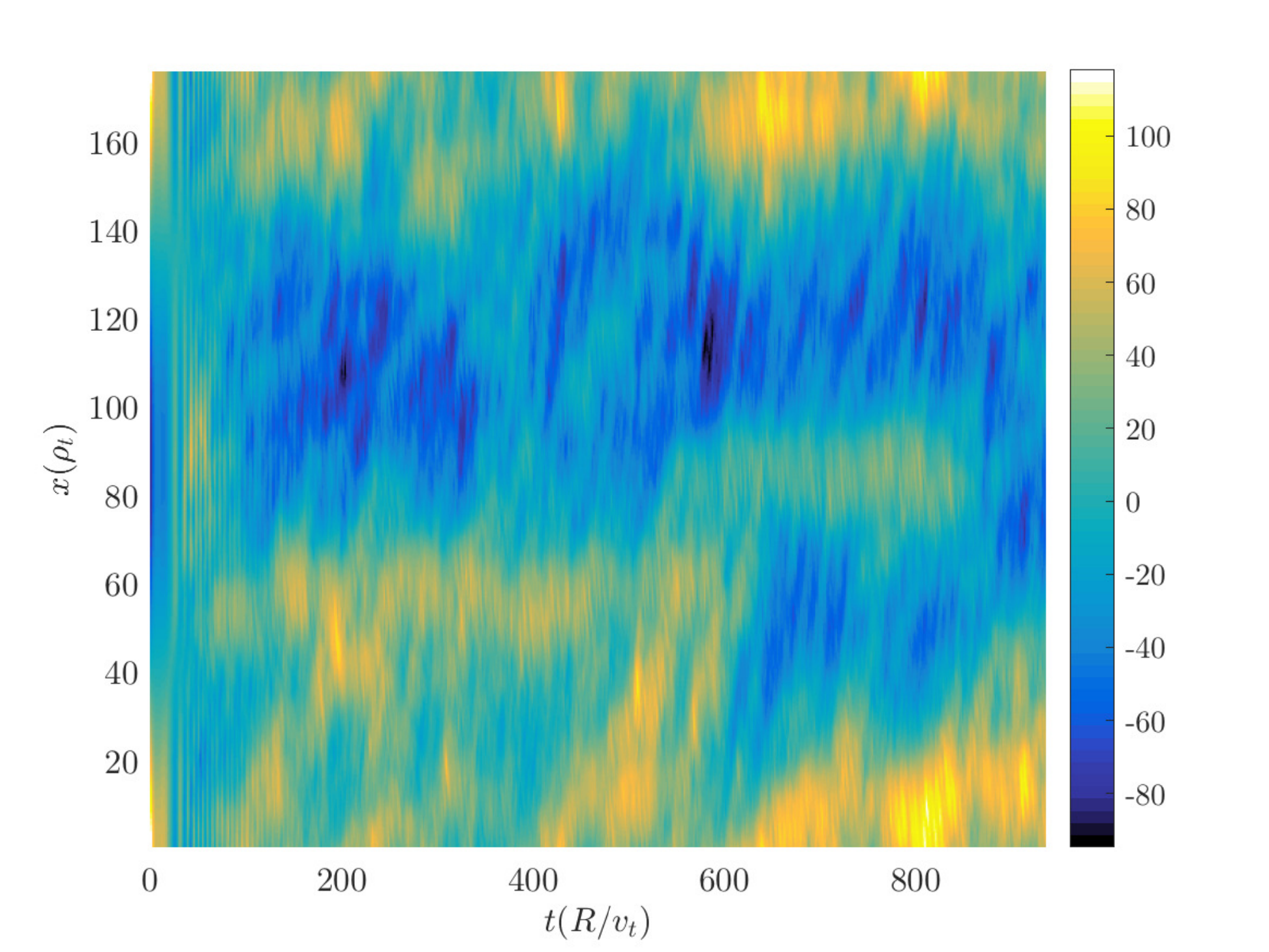} \caption{} \label{fig:zonal_corrected_cyc_itg2}
\end{subfigure}

\begin{subfigure}{0.48\textwidth}
\includegraphics[width=8.5cm]{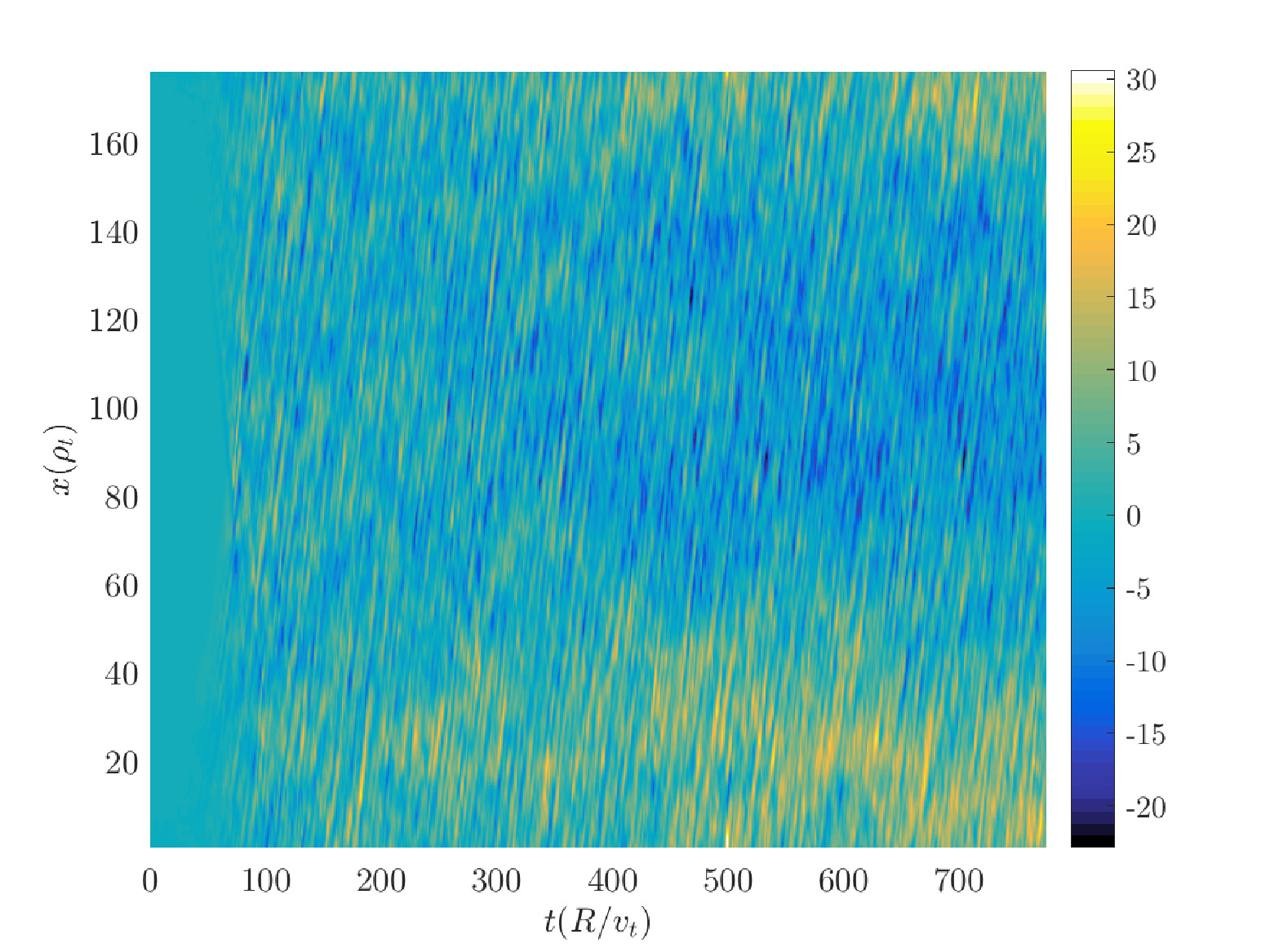} \caption{} \label{fig:zonal_original_cyc_etg}
\end{subfigure}
\begin{subfigure}{0.48\textwidth}
\includegraphics[width=8.5cm]{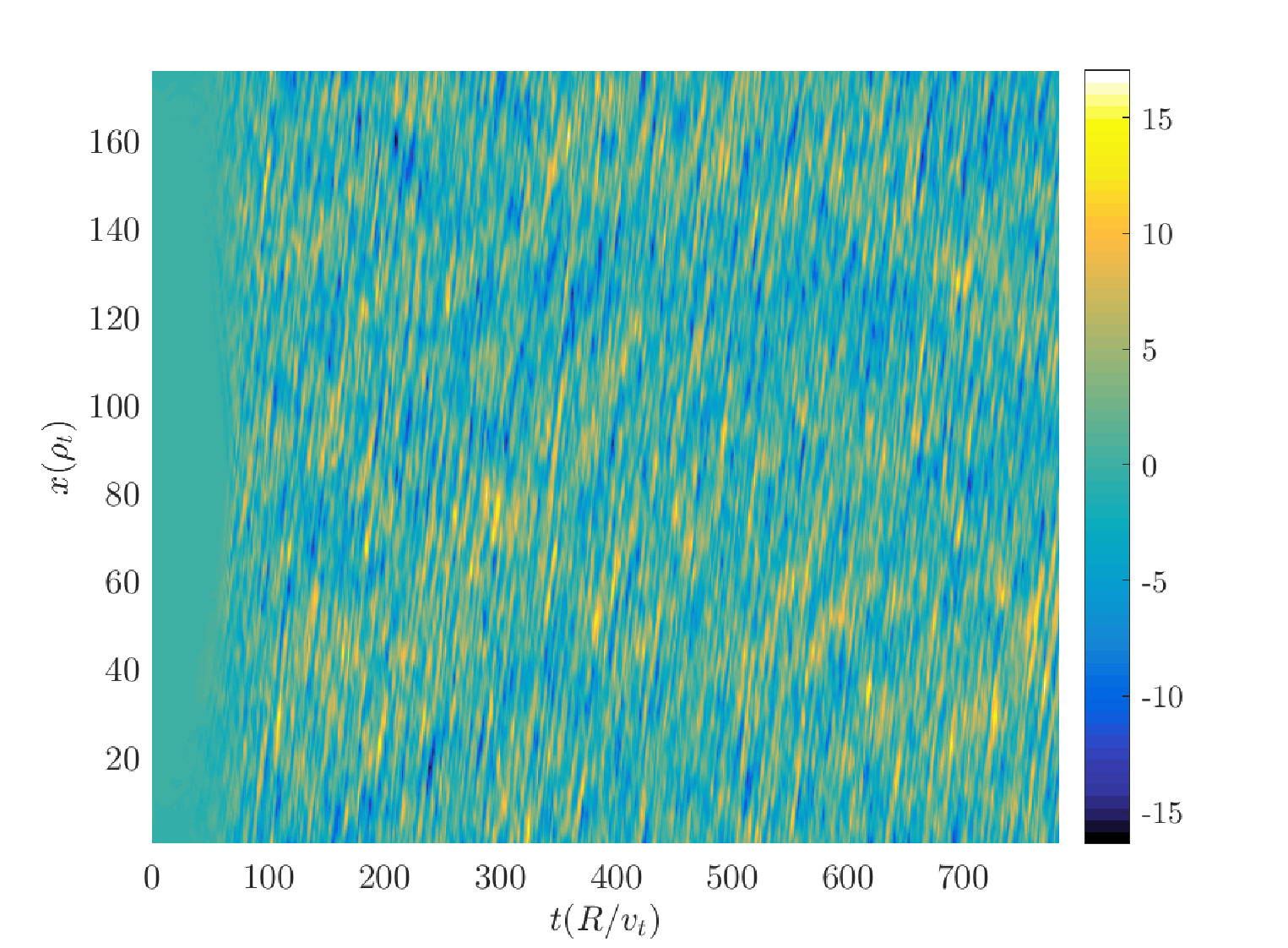} \caption{} \label{fig:zonal_corrected_cyc_etg}
\end{subfigure}

\caption{Zonal potential in Cyclone case with $S=0.12 v_t/R$, (a) ITG original remap (b) ITG corrected remap, (c) ETG original remap and (d) ETG corrected remap.
}
\label{fig:zonal_cyc}
\end{figure}

\begin{figure}[htb]
\centering
\epsfig{figure=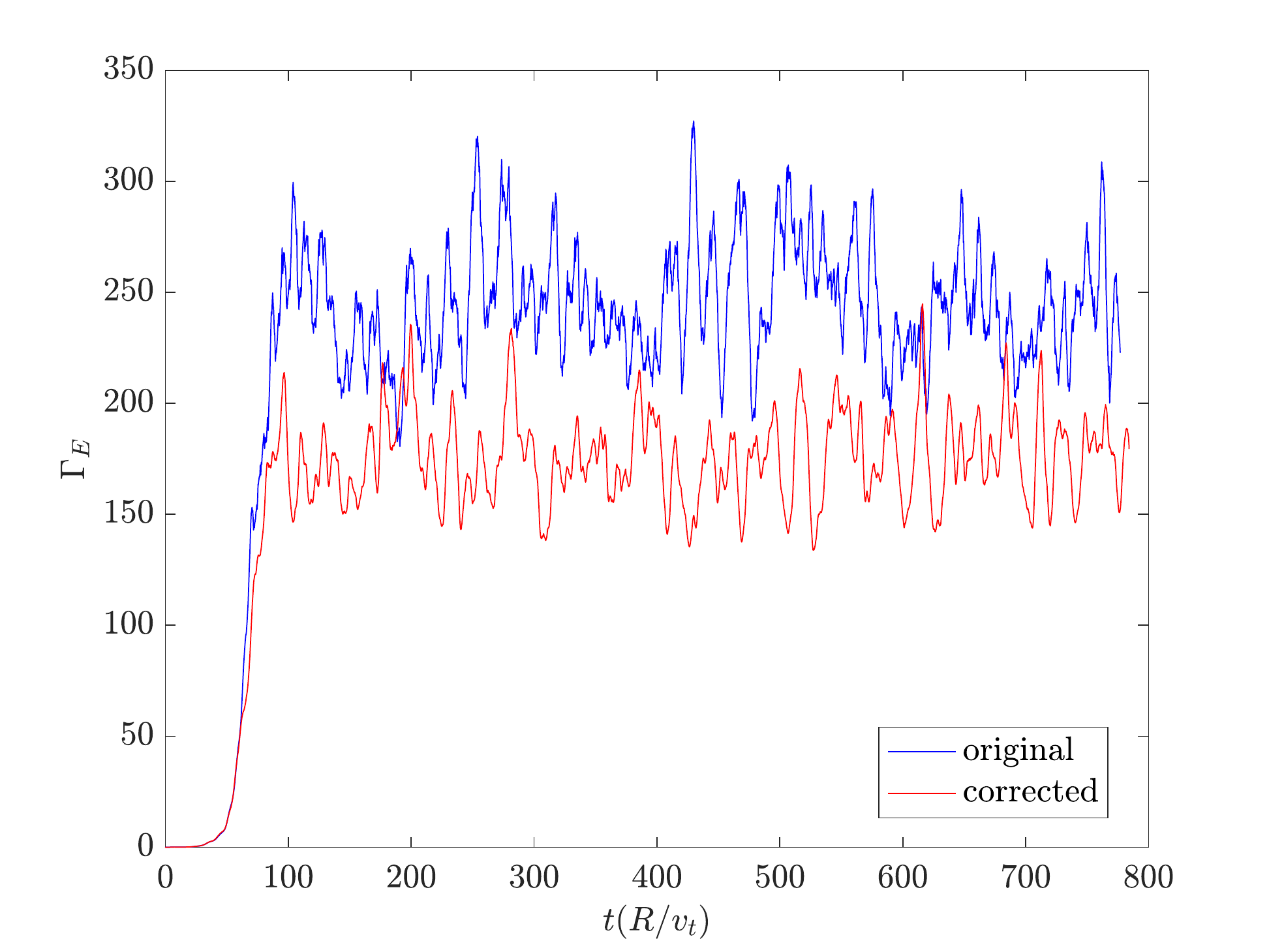,width=11cm}
\caption{Heat flux (in units of $n m v_t^3 \rho_t/R$) versus time in CYCLONE ETG case with $S=0.12 v_t/R$.}
  \label{fig:heatflux_cyclone_ETG}
\end{figure}

The ITG, but not ETG, saturation phyics is essentially dominated by zonal flows (see figure \ref{fig:zonal_cyc}). Zonal flow structures are somewhat different between the original and corrected remap simulations; ITG simulations do not exhibit obvious systematic differences, but in the ETG simulations, there is a significantly larger zonal flow observed at the system scale in the original rather than corrected remap simulation. The original and corrected remap simulations have similar fluxes in the ITG case but in the ETG case the corrected-remap fluxes are $\sim 30\%$ lower (see figures \ref{fig:heatflux_cyclone_ETG} and \ref{fig:heatflux_cyclone}). There are stronger long-wavelength zonal flows, however, in the original-remap ETG case.

\subsection{Waltz standard case}

Waltz standard case parameters are $q=2$, shear $(dq/dr)r/q = 1$, density and temperature gradient scale lengths $L_n = R/3$ and $L_t = R/9$ respectively, local aspect ratio $a/R=1/3$, and equal ion and electron temperatures, and the simulation is run in concentric circular geometry.

Two sets of simulation sizes were run, one with a standard box size ($L_x=180\rho_{t}$, $L_y = 125\rho_{t}$) and one with a box that was twice as large ($L_x = 360\rho_{t}$, $L_y =250\rho_{t}$): the standard box size is typical of a reasonably well resolved simulation, and the larger simulation box is unusually large.

\begin{figure}[htb]
\centering
\epsfig{figure=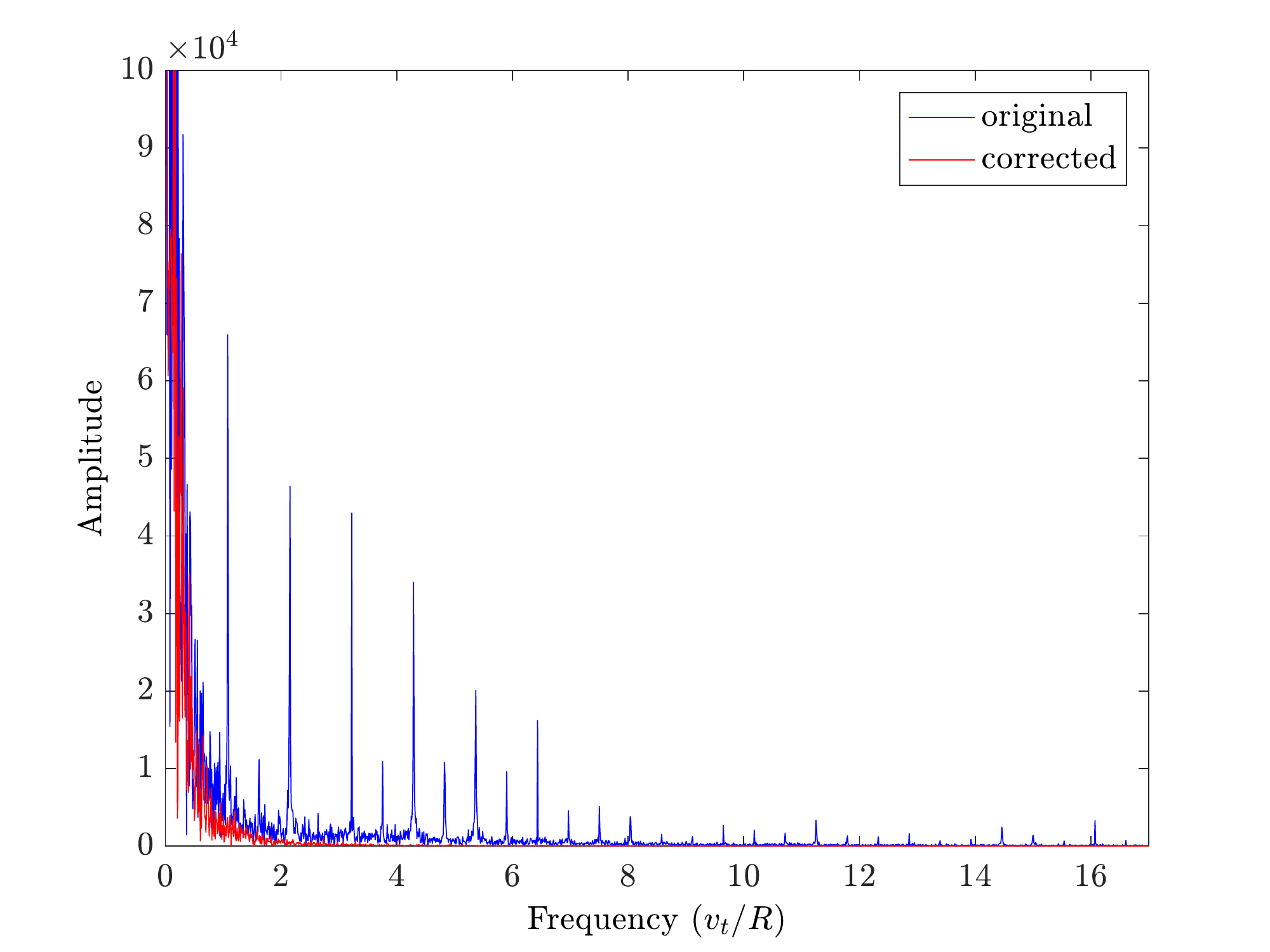,width=11cm}
\caption{Absolute value of temporal Fourier transform of volume-averaged heat flux in the large Waltz ETG case.}
\label{fig:fluxfft}
\end{figure}

\begin{figure}[htb]
\centering
\epsfig{figure=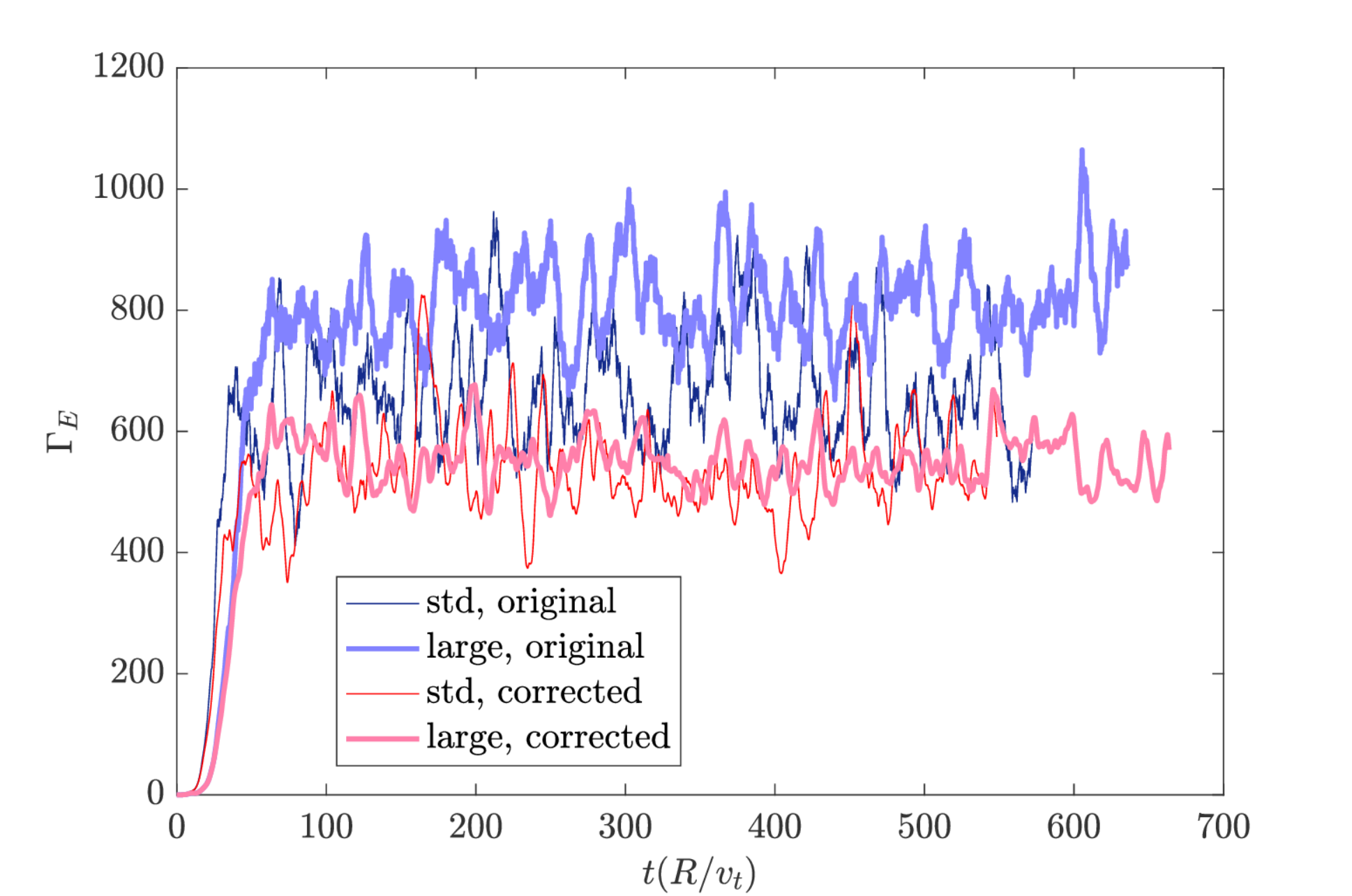,width=11cm}
\caption{Heat flux (in units of $n m v_t^3 \rho_t/R$) versus time in Waltz ETG case: thin traces are standard simulation domain and thick traces are double domain. Blue traces are original remap and red traces are corrected remap.}
\label{fig:waltz_etg_flux}
\end{figure}

For these ETG cases, the flux traces are much noisier for original rather than corrected remap (figure \ref{fig:waltz_etg_flux}). Note that these simulations again have $S=0.12 v_t/R$, where the thermal velocity is that of the electrons: this is exceptionally large compared to experimental values. There are periodic downwards dips to the flux for the large original-remap case. The corrected remap traces on the other hand look quite smooth. The apparently unphysical jumps in the flux are seen more clearly in the temporal Fourier transform of this quantity (figure \ref{fig:fluxfft}), where the harmonics of the remap frequency (in angular frequency units, this is $\omega_R = 2 \pi S k_{y0}/ k_{x0} = 1.09 v_t/R$) are seen as large spikes in the spectrum.

The fluxes for the large original remap case are around 20\% larger than for the standard size case. Fluxes for the two corrected-remap simulations are somwhat different, so do not appear to have converged with size: since the difference between the original- and corrected-remap simulations increases for the larger case, the original-remap fluxes do not appear to be converging to the value of the corrected-remap simulations. In other words, the error due to incorrect nonlinear coupling does not reduce as the simulation box size is increased.

\begin{figure}
\begin{subfigure}{0.48\textwidth}
\includegraphics[width=8.5cm]{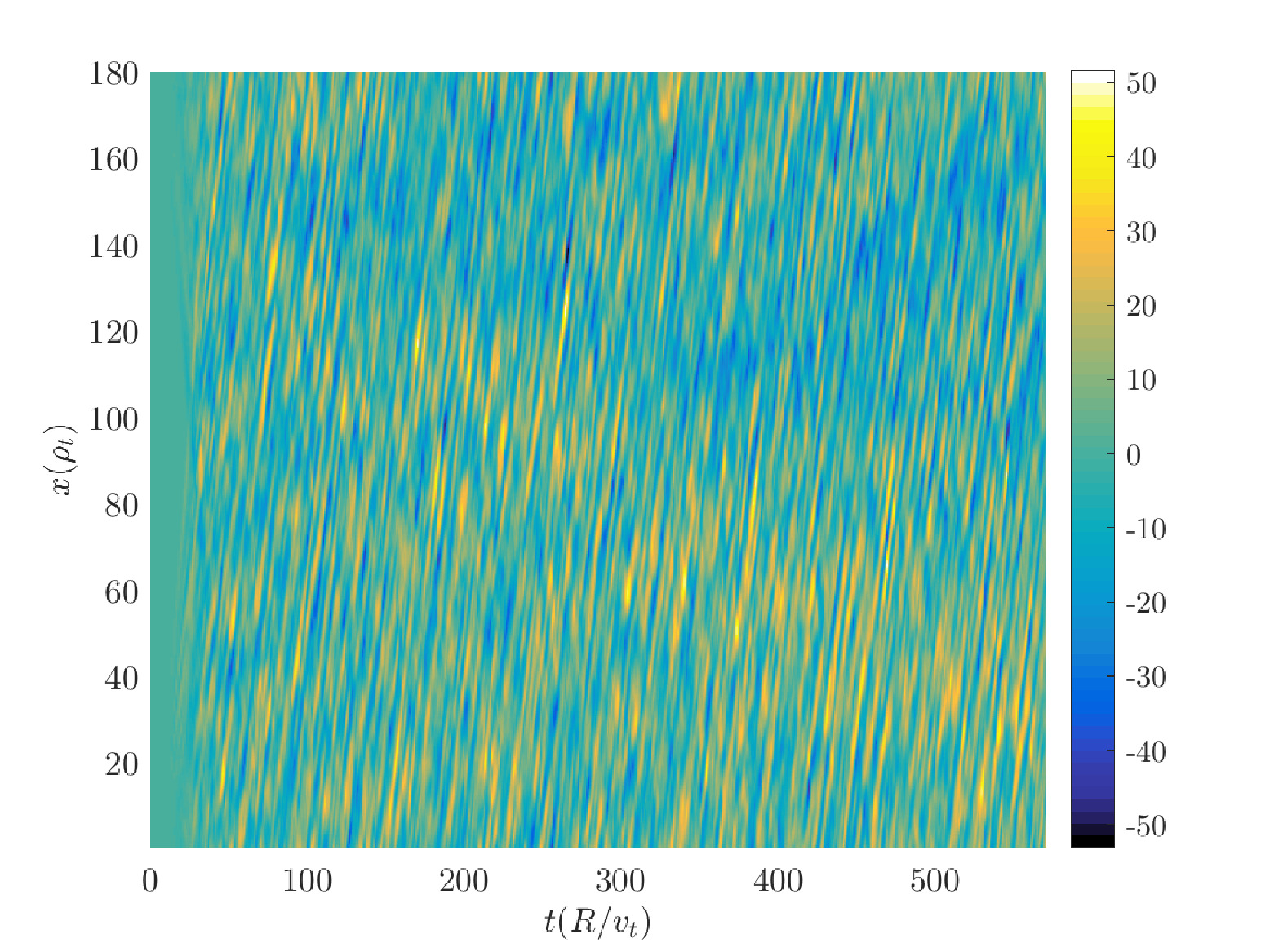} \caption{} \label{fig:zonal_original_waltz_etg_std}
\end{subfigure}
\begin{subfigure}{0.48\textwidth}
\includegraphics[width=8.5cm]{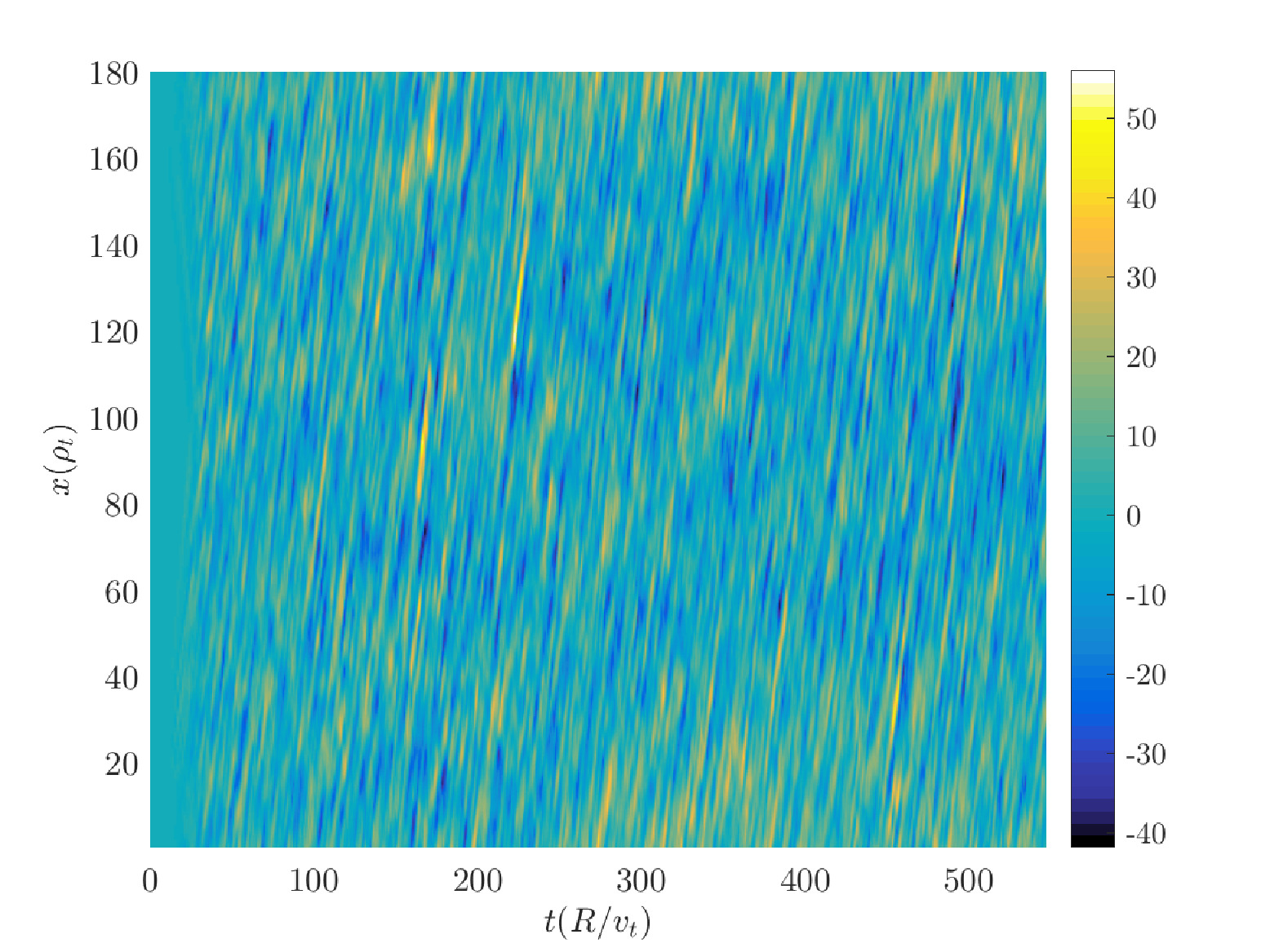} \caption{} \label{fig:zonal_corrected_waltz_etg_std}
\end{subfigure}

\caption{ Zonal potential in Waltz ETG standard size case, (a) original remap and (b) corrected remap.}
\label{fig:zonal_waltz_etg}
\end{figure}

The Waltz ETG cases have very weak zonal flows compared to typical ITG simulations (figure. \ref{fig:zonal_waltz_etg}), and there is no clear systematic difference between the flows observed in original and corrected remap simulations.

\begin{figure}
\begin{subfigure}{0.48\textwidth}
\includegraphics[width=8.5cm]{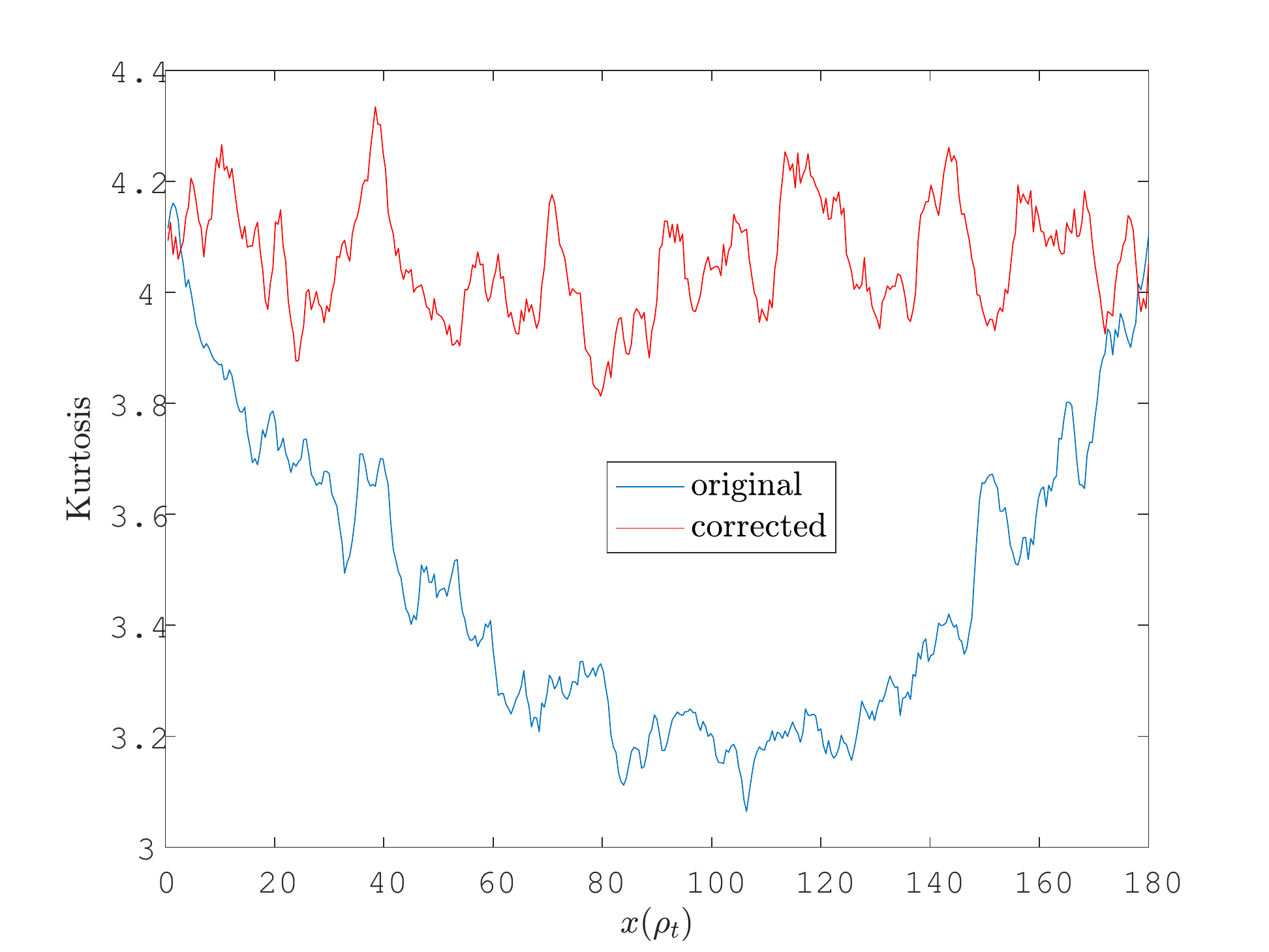} \caption{} \label{fig:kurtosis_waltz_standard}
\end{subfigure}
\begin{subfigure}{0.48\textwidth}
\includegraphics[width=8.5cm]{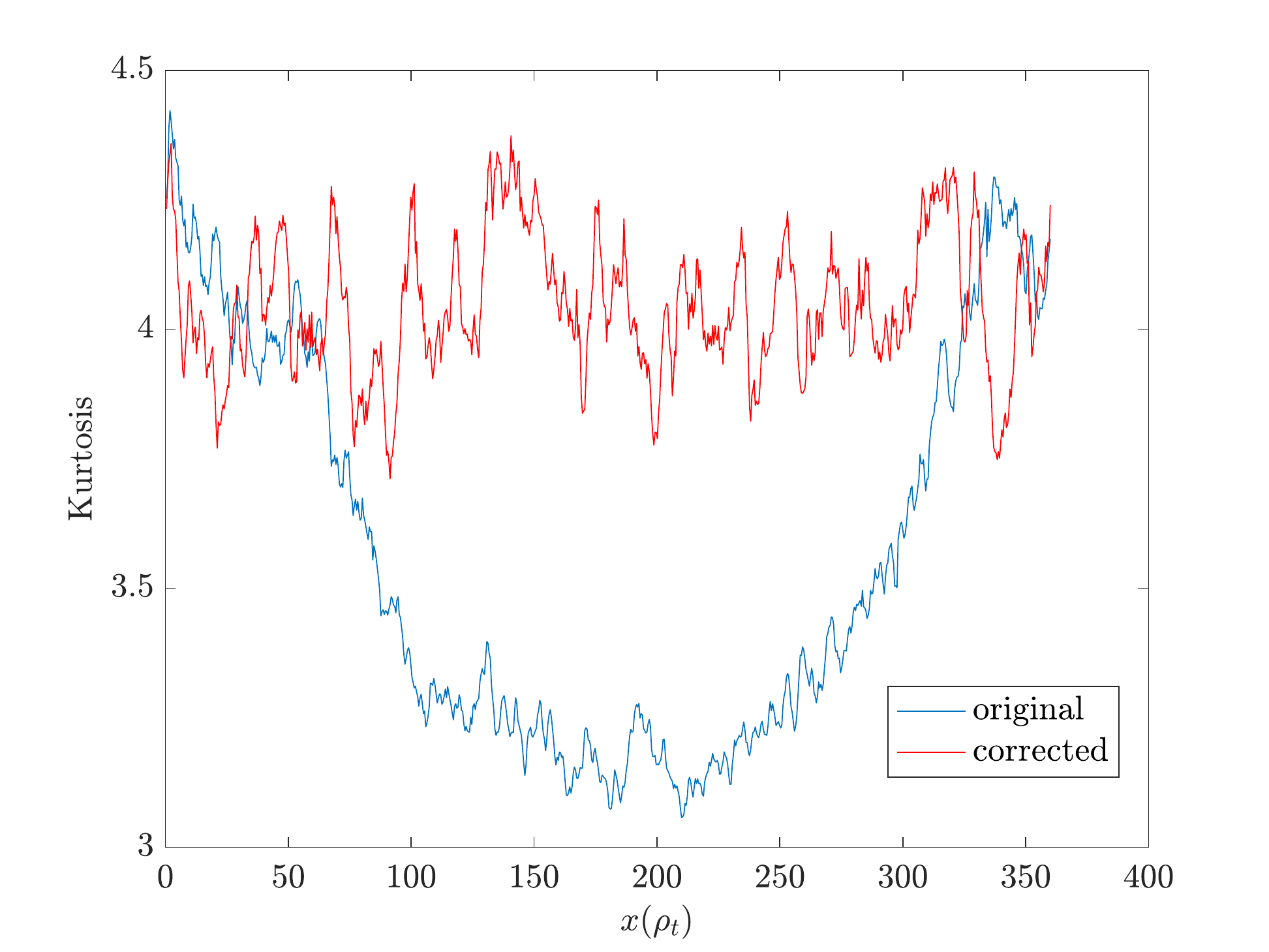} \caption{} \label{fig:kurtosis_waltz_large}
\end{subfigure}

\caption{ Kurtosis versus radius (last half of simulation time domain) in Waltz ETG case with (a) standard and (b) doubled simulation domain.}
\label{fig:kurtosis_waltz}
\end{figure}

Figure \ref{fig:kurtosis_waltz} shows that, except at the edges of the simulation domain, the kurtosis is around 3 for both the standard and large uncorrected remap simulations, suggesting that the wave phases aren't correlated. The edges (i.e. near $x=0$ and $x=1$) are special in original-remap because the individual plane waves `jump' little there during a remap operation. Higher kurtosis (around 4) is seen for corrected remap, at at the edges of the domain in standard remap, indicating that the turbulence is somewhat intermittent. The kurtosis in the corrected remap simulation tends to be relatively uniform, and around 4. Taking into account the magnetic shear, the simulation equations have a nine-fold discrete translation symmetry in the radial direction, so we would not expect to see variations in the statistical properties of the turbulence on the box scale; the corrected-remap simulation is consistent with the underlying symmetry of the system, but the original-remap simulation is not. Note that this breaking of symmetry in original-remap does not appear to reduce when the simulation size increases.

\section{Conclusions}

We have presented a method, that we call the corrected wavevector-remap, for implementing homogeneous background shear flows in gyrokinetic systems using a spectral representation on the plane perpendicular to the field line. For the gyrokinetic problem, this is relatively straightforward to implement, and there is essentially no computational penalty to using a corrected wavevector-remap method compared to the original wavevector-remap method. This method is generally applicable to spectral representations of problems in shearing box geometry, so should also be suitable in certain problems arising elsewhere, for example in accretion disk physics. 

In the astrophysical context the remap method\cite{Umurhan2004} refers to a moving grid method, with infrequent remappings from the sheared Lagrangian frame to a orthogonal coordinate system, at time points where every mode has been advected by an integer number of grid steps. This is quite closely related to the method presented here, although instead of infrequently making a large change to the region of wave-space under consideration, the wave-space in the corrected wavevector-remap method stays roughly rectangular. 

Although the original wavevector-remap\cite{HammettRemap} method was motivated fairly clearly in terms of the effect of background flow on a wave spectrum, this justification was not completely systematic, and it was unclear whether results based on this method were trustworthy. Previously, the use of linear test cases and comparisons with non-periodic codes have been used as support for the appropriateness of the original-remap. Linear results\cite{CassonPHD} show convergence in the limit $L_x\rightarrow \infty$, and converge sufficiently rapidly that typical nonlinear simulation boxes are sufficiently large to resolve the linear physics. On the other hand, certain worrying features including the sudden jumps in real space diagnostics suggested that there were unresolved issues. By relating the corrected remap method to the original remap method, we were able to clarify what aspects of original remap lead to errors, and, in particular, described the off-by-one errors in nonlinear coupling in original remap.

A simple test problem, finding the nonlinear coupling resulting from two large-amplitude test waves subject to a background shear flow, has been solved both analytically and using the original and corrected remap schemes. The corrected remap scheme agrees well with the analytical solution, supporting the claim that the formulation and implementation in the GENE code are correct. On the other hand, the rather poor behaviour of the original remap method, which does not converge to the correct result in the large system-size limit calls into question the appropriateness of original remap for nonlinear physics simulations. 

In ITG simulations, where zonal flows play a dominant role in turbulence saturation, the original and corrected remap simulations had heat and momentum flux levels that were broadly comparable. Details of the turbulence properties, however, are quite different, and there is a strong radial variation in turbulence intensity and intermittency in the original-remap simulation that is not consistent with the expected symmetry. ETG test case simulations, where saturation is not dominated by zonal flows, show more significant differences in flux levels between the original remap and corrected remap methods. This is consistent with analysis suggesting that the nonlinear coupling between a zonal and an $n \neq 0$ mode is correctly treated by the original remap.

Overall, the heat flux levels in Waltz standard case ETG simulations were somewhat affected by the remap error, but perhaps more significantly there were substantial changes in qualitative features of the turbulence. The kurtosis statistic, which measures how long-tailed the distribution of turbulence intensity is, shows that the statistical features of the corrected remap simulations are quite different to those of the original remap simulations. It also shows that original remap simulations have inhomegeneous turbulence properties, with longer-tailed fluctuation amplitude near the simulation boundary in the $x$ direction, artificially breaking the symmetry of the gyrokinetic model. We might expect simulations with very intermittent fluxes (for example, for edge plasmas) to potentially be quite strongly affected by the errors in original remap. Note also that the original-remap simulations did not appear to converge when simulation-size was increased, unlike the corrected-remap simulations.

Overall, despite the original remap scheme performing reasonably well in many test cases, it seems preferable for future spectral simulations with a background shear flow to instead use a corrected remap method. The required changes to the nonlinear term are straightforward to implement and have an insignificant computational cost. Moving to the corrected method seems be particularly desirable where localised structures arise in the binormal direction\cite{wyk17}, which are likely to be particularly sensitive to artificial jumps in wave phases. Also, it would be desirable to re-run and re-analyse the classic test cases using the corrected-remap method and compare this carefully against approaches such as that of ref. \citet{candy_remap} or ref. \citet{McMillan2015} that do not use periodic boundaries. 

\acknowledgements{
 We would like to acknowledge useful conversations with F. Casson and C. Roach. This work has been carried out within the framework of the EUROfusion Consortium and has received funding from the Euratom research and training programme 2014-2018 under grant agreement No 633053. The views and opinions expressed herein do not necessarily reflect those of the European Commission. This work is partly funded under EPSRC EP/N035178/1 and EP/R034737/1. Simulations were performed with the support of Eurofusion and MARCONI-Fusion. This work was supported by a grant from the Swiss National Supercomputing Centre (CSCS) under project ID s793. This work was supported in part by the Swiss National Science Foundation.
}


\end{document}